\documentclass[%
onecolumn,
 amsmath,amssymb,
 aps,
 prd,
]{revtex4-2}

\usepackage{graphicx}
\usepackage{dcolumn}
\usepackage{bm}
\usepackage{hyperref}
\usepackage[dvipsnames]{xcolor}

\newcommand{\der}{\mathrm{d}}

\begin{document}

\title{Gauge-invariant perturbations of relativistic non-perfect fluids in spherical spacetime}

\author{David Díaz-Guerra}
\email{ddiazgue@usal.es}
\author{Conrado Albertus}
\email{albertus@usal.es}
\author{Prasanta Char}
\email{prasanta.char@usal.es}
\author{M.Ángeles Pérez-García}
\email{mperezga@usal.es}
\affiliation{Department of Fundamental Physics and IUFFyM, University of Salamanca, Plaza de la Merced S/N E-37008, Salamanca, Spain}


\begin{abstract}
Astrophysical compact objects are usually studied using a perfect fluid model.
However, in astrophysical processes out-of-equilibrium, dissipative effects become important to describe the dynamics of the system.
In this work we obtain gauge-invariant non-spherical perturbations of a self-gravitating non-perfect fluid in spherical spacetime.
We use the Gerlach-Sengupta formalism to work with gauge-invariant metric perturbations, and the Gundlach-Martín-García approach to transform the equations of tensor perturbations into scalar equations.
We calculate the dynamics of the dissipative contributions, e.g., viscosity and heat flux, using the Müller-Israel-Stewart equations in the gauge-invariant formalism.
We obtain a set of field equations for the evolution of matter and metric perturbations in the polar and axial sectors. 
Specifically, in the former we find two wave equations sourced by the anisotropic contributions, and the evolution of all matter perturbations for radiative modes ($l\geq 2$).
In the axial sector, we find one wave equation coupled to the evolution of matter perturbations. 
Finally, we comment on the contribution of dissipative effects in the lower-order multipoles ($l=0,1$) for both sectors. 
\end{abstract}

\maketitle

\section{\label{sec:intro}Introduction}

Neutron Stars (NSs) are nearly spherical compact objects born hot and dense. In particular, the internal temperature at the time of their birth may reach $\sim 8 \times 10^{10}$ K, according to core-collapse supernova simulations \cite{Janka:2006fh}. 
In addition, binary neutron star (BNS) mergers may also lead to the formation of hypermassive neutron stars with similar initial high temperature \cite{Baiotti:2016qnr}. 
The observational characteristics of NSs, such as their oscillation spectra and rotation frequency, are dependent not only on their composition, namely the Equation of State (EoS), but also on the transport properties of matter inside them. 
For example, the bulk viscosity provides a damping influence on the oscillation modes, which would have been otherwise unstable \cite{Cutler1987}. 
In this way, the dissipative properties of matter control the dynamical stability of the aforementioned dense stars. 

The effect of dissipative processes on the stability of the stellar models has been discussed in the literature \cite{Hiscock:1983zz,Lindblom1983,Cutler1987}. The transport properties of neutron star matter have also been investigated with detailed microphysical inputs in recent years \cite{Shternin:2008es,Manuel:2012rd,Shternin:2013pya,Alford:2017rxf} even including dark exotic components \cite{ANGELESPEREZGARCIA2022136937}. In General Relativity (GR), dissipation was treated by first order expansions of the gradients of hydrodynamic variables of a perfect fluid by Eckart \cite{Eckart:1940te}, Landau and Lifshitz \cite{landau2014}, and recently by Bemfica et al. \cite{Bemfica:2020zjp}. The first order theories are plagued by the issues of lack of causality and stability.  Another approach was pioneered by Müller, Israel, and Stewart (MIS) \cite{Muller:1967zza,Israel:1979wp}, known as the extended irreversible thermodynamics, which contains terms that are of the second order in the dissipative thermodynamic fluxes, such as the viscosity and heat flux. 
The situation becomes more complicated when we consider {the dissipative effects on the fluid perturbations} in order to understand the gravitational wave (GW) emission from these objects. 

In this work, we choose the gauge-invariant formalism to describe a self-gravitating {perturbed} spherical system, first proposed by Gerlach and Sengupta (GS) \cite{Gerlach:1979rw, Gerlach:1979ih, Gerlach:1980tx}. This formalism was further developed by Gundlach and Mart{\'i}n-Garc{\'i}a (GM), using scalar equations for perfect fluids \cite{Gundlach:1999bt,Martin-Garcia:2000cgm}. 
We use this formalism to derive the equations from a non-perfect fluid, including dissipation viscosity and heat flow. The gauge-invariant formalism has been widely used to study the GWs from a spherical collapse to black hole (BH) \cite{Seidel:1990xb, Harada:2003bu}, and accretion-driven gravitational radiation \cite{Nagar:2004ns}. It has also been applied to study the coupling between radial and non-radial oscillation modes in compact stars \cite{Passamonti:2004je,Passamonti:2005cz,Passamonti:2007tm}. For a review of the gauge-invariant treatment of perturbed Schwarzschild spacetime, see \cite{Martel:2005ir,Nagar:2005ea}. In this work, we use this formalism, for the first time, to study the characteristics of a perturbed non-perfect fluid and its equations of motion.

This paper is organized as follows. 
In section~ \ref{sec:fluid}, we review the relativistic non-perfect fluid, using extended irreversible thermodynamics to describe its dynamical evolution. 
In section~ \ref{sec:GS_formalism}, we review the GS formalism of non-spherical perturbations over a spherical background. We describe the time-radial and angular decomposition of the dynamical background equations to perturb around this configuration. 
In section~ \ref{sec:pert_fluid}, we split the description of the non-perfect fluid into background plus perturbation parts, and we apply the GM formalism to work with the perturbations as scalars. 
In section~ \ref{sec:evolution}, we calculate the field equations for non-spherical perturbations of metric and matter of a non-perfect fluid. Finally, in section~ \ref{conclude}, we {conclude and} discuss future directions regarding the lines exposed in this work.  
We use {geometrized units by} setting $c =G =k_B = 1$. 

\section{Review of the relativistic non-perfect fluid dynamics} \label{sec:fluid}
We work with a self-gravitating relativistic fluid set in a dynamical spacetime that is described by a 4-dimensional manifold $\mathcal{M}$. The coordinates in this space are $x^\mu$($\mu =~0,1,2,3$). This manifold has a metric tensor $\mathbf{g}$, and a covariant derivative $\tilde \nabla$ that defines a connection $^{(4)}\Gamma^{\mu}_{\nu \rho}$ to the partial derivatives of the coordinates $x^\mu$.

Some tensor quantities on $\mathcal{M}$, e.g., the fluid velocity $\mathbf{u}$ or the stress tensor $\mathbf{T}$, describe physical magnitudes of the fluid, as we will show below. The self-gravitating fluid is in a dynamical spacetime, therefore: (i) the conservation relations of such tensors characterize the evolution of the fluid, and (ii) the stress tensor of the fluid $\mathbf{T}$ is the source of the Einstein field equations (EFE) that governs the dynamics of the spacetime. In this section, we detail the first item (i), summarizing the hydrodynamics of a relativistic non-perfect fluid.

\subsection{Fluid thermodynamics}
A subset of the tensor quantities that define the fluid are scalar thermodynamic variables. We define the local matter number density, $n$, energy density, $\varepsilon$ and the specific entropy $s$, as their correspondent magnitudes measured by an observer comoving with the fluid. These scalar quantities follow the fundamental law of thermodynamics in equilibrium,
\begin{equation}
    \der \varepsilon  =  \frac{(\varepsilon + p)}{n} \der n + n T \der s. \label{eq:fund_law}
\end{equation}
From this relation, the temperature $T$ and the isotropic pressure $p$ of the fluid are
\begin{align}
    T &\equiv \frac{1}{n} \left( \frac{\partial \varepsilon}{\partial s}\right)_n, \label{eq:def_T} \\ 
    p &\equiv n \left( \frac{\partial \varepsilon}{\partial n}\right)_s - \varepsilon. \label{eq:def_p}
\end{align}
Likewise, the microphysics of the system imposes an Equation of State (EoS), $\varepsilon = \varepsilon (n,s)$, that relates these quantities. Therefore,  Eq. \eqref{eq:fund_law}, and the EoS of the system relate all the thermodynamic variables and, as a consequence, only two equilibrium thermodynamic quantities result independent.

\subsection{Generic relativistic fluid hydrodynamics} 

The main tensor quantities that describe the dynamics of a generic fluid are the density current $\mathbf{J}$, and the stress-energy tensor $\mathbf{T}$. From a timelike vector $u_\mu$, describing the fluid velocity we decompose the current and the stress tensor as in \cite{Israel:1979wp},
\begin{subequations}
    \label{eq:current_and_stress}
  \begin{align}
    J_{\mu} &= nu_{\mu} + \mathcal{J}_{\mu}, \\
    T_{\mu \nu} &= \varepsilon u_{\mu} u_{\nu} + \mathcal{P} \Delta_{\mu \nu} + 2 \mathcal{Q}_{(\mu}u_{\nu)} + \mathcal{T}_{\mu \nu}, \label{eq:general_stress}
\end{align}
\end{subequations}
where $\Delta_{\mu \nu} \equiv g_{\mu \nu} + u_\mu u_\nu$ is the {orthogonal} projection tensor. $n \equiv - u^{\mu}J_{\mu}$, $\varepsilon \equiv  u^\mu u^\nu T_{\mu \nu} $, and $\mathcal{P} \equiv \frac{1}{3}\Delta^{\mu \nu} T_{\mu \nu}$ are scalars. The vectors $\mathcal{J}_{\mu} \equiv \Delta_{\mu}^{ \nu} J_{\nu}$ and $\mathcal{Q}_{\mu} \equiv - u^{\alpha} T_{\mu \nu} \Delta^{\nu}_\alpha$ are transverse to $u_\mu$. 
The tensor $\mathcal{T}_{\mu \nu} \equiv \frac{1}{2} \left( \Delta_{\mu}^\alpha \Delta_{\nu}^\beta + \Delta_{\mu}^\beta \Delta_{\nu}^\alpha - \frac{2}{3} \Delta^{\alpha \beta}_{\mu \nu} \right)$ is transverse to $u_\mu$, symmetric and traceless. This decomposition derives from the irreducible representations of a vector and a symmetric tensor using a generic timelike vector, $u_\mu$ while for general fluids, the local rest frame of the observer comoving with the fluid velocity is ambiguous however. 

We choose the Eckart frame, in which the fluid velocity is defined as parallel to the current flux $\mathbf{J}$, by setting $\mathcal{J}_\mu = 0$. Therefore, the number density $n$ and energy density $\varepsilon$ are the thermodynamic variables appearing in the fundamental law of thermodynamics in Eq. \eqref{eq:fund_law}.

The conservation equations on the current $\mathbf{J}$, and stress tensor $\mathbf{T}$, describe the dynamics of the fluid,
\begin{align}
    \tilde \nabla^\nu J_{\nu} &=0, \label{eq:current_cons}\\
    \tilde \nabla^\nu T_{\mu \nu} &= 0. \label{eq:stress_cons}
\end{align}
Eq. \eqref{eq:current_cons} and the projection of $u^\mu$ and $\Delta^{\mu \nu}$ into Eq.~\eqref{eq:stress_cons}, yield the matter and energy conservation along with  Euler equations
\begin{subequations}
    \label{eq:conservation_laws}
    \begin{align}
        u^\mu \tilde \nabla_\mu n + \Theta n &=0, \label{eq:cons_matter} \\
        u^\mu \tilde \nabla_\mu \varepsilon + \Theta \left( \varepsilon + \mathcal{P}\right) + \sigma^{\mu \nu} \mathcal{T}_{\mu \nu}&\nonumber \\ 
        + 2 \mathcal{Q}_\mu a^\mu + \mathcal{D}^\mu \mathcal{Q}_\mu  &= 0, \label{eq:cons_energy} \\
        \left( \varepsilon + \mathcal{P}\right) a_\mu + \mathcal{D}_{\mu} \mathcal{P} + \mathcal{D}_\nu \mathcal{T}^{\nu}_\mu + \Delta ^\nu _\mu u^\lambda \tilde \nabla_\lambda \mathcal{Q}_\nu & \nonumber \\ 
        + \left(\omega_{\mu \nu} + \sigma_{\mu \nu} + \frac{4}{3}\Theta\Delta_{\mu \nu}   \right) \mathcal{Q}^\nu &= 0,  \label{eq:cons_euler}
    \end{align} 
\end{subequations}
with the expansion scalar $\Theta$, fluid acceleration $a_\mu$, orthogonal derivative $\mathcal{D}_\mu$, shear tensor $\sigma_{\mu \nu}$, and vorticity tensor $\omega_{\mu \nu}$, defined as follows
\begin{subequations}
 \begin{eqnarray}
    \Theta &\equiv& \tilde \nabla_{\mu} u^{\mu}, \\
    a^{\mu} &\equiv& u^\nu \tilde \nabla_\nu u^\mu , \\
    \mathcal{D}_\mu X_{\nu \dots \lambda} &\equiv& \Delta^\alpha_\mu \Delta^\beta_\nu \dots \Delta_\lambda^{\gamma} \tilde \nabla_\alpha X_{\beta \dots \gamma}, \\
     \sigma_{\mu \nu} &\equiv& \tilde \nabla_{(\mu} u_{\nu)} + a_{(\mu}u_{\nu)} - \frac{1}{3}\Theta \Delta_{\mu \nu},\\ 
    \omega_{\mu \nu} &\equiv& \Delta^\alpha_\mu \Delta^\beta_\nu \tilde \nabla_{[\alpha} u_{\beta ]}.
\end{eqnarray}
\end{subequations}
We use the notation of parenthesis, $X_{(\mu \nu)} \equiv\frac{1}{2} \left( X_{\mu \nu} + X_{\nu \mu} \right)$ for symmetrization of indices, and brackets, $X_{[\mu \nu]} \equiv\frac{1}{2} \left( X_{\mu \nu} - X_{\nu \mu} \right)$ for anti-symmetrization of indices.

The terms $u^\mu \tilde \nabla_\mu X$ in Eqs. \eqref{eq:conservation_laws} represent the evolution or ``rate of change'', that an observer comoving with the fluid, i.e. following the worldines of $u_\mu$, is measuring on a given quantity $X$.

We have described the evolution of the number density $n$ in Eq. \eqref{eq:cons_matter}, the energy density $\varepsilon$ in Eq. \eqref{eq:cons_energy}, and the components of the vector $\mathcal{Q}_\mu$ aligned to the fluid movement in Eq. \eqref{eq:cons_euler}. However, {the evolution of the rest of components} of $\mathcal{Q}_\mu$, and the tensor $\mathcal{T}_{\mu \nu}$ are not contained in these equations. For a perfect fluid, the quantities $\mathcal{Q}_{\mu}$ and $\mathcal{T}_{\mu \nu}$ are zero. For a non-perfect fluid, these quantities do not vanish and we need additional evolution equations. 

\subsection{Non-perfect fluid description}
A non-perfect fluid deviates from the perfect fluid approximation by including the effects of dissipation. 
The dissipative effects are a consequence of irreversible processes that increase the entropy of the system. Therefore, the dissipative quantities are not present in the equilibrium state. Then, on theoretical grounds we can split relevant quantities of the fluid into contributions that exist in equilibrium plus dissipative terms, that vanish in such state. 

As already discussed, this is relevant for a realistic description of the interior of NSs. This viscous character of real fluids has been mostly neglected from calculations in the literature due to complexity. It is well known that no real liquid exhibits zero viscosity. As an example, for all weakly coupled theories in the dilute limit there is a lower bound which one can derive from the uncertainty principle i.e. the ratio of shear viscosity to entropy density $\eta/s > 1/12$. For certain strongly coupled theories a lower bound  from the AdS/CFT conjecture  $\eta/s > 1/4\pi$ \cite{CREMONINI2011}.

A relativistic fluid is in equilibrium if its inverse-temperature vector $\beta_{\mu}$, is a Killing vector field \cite{Israel:1979wp},
\begin{align}
    \tilde \nabla_{(\mu}\beta_{\nu)} = 0, \label{eq:equil_beta}
\end{align}
with $\beta_\mu \equiv \frac{u_\mu}{T}$.

In the case of self-gravitating fluid, this condition is equivalent to the spacetime being stationary, i.e., there exists a time-like Killing vector. In the equilibrium state, the following conditions are satisfied
\begin{subequations}
    \label{eq:equilibrium_conds}
    \begin{align}
        \Theta &= 0, \\ 
        \sigma_{\mu \nu} &= 0, \\ 
        \mathcal{D}_{\mu} \ln T + a_{\mu} &= 0,
    \end{align}
\end{subequations}
which indicate that the fluid does not exhibit scalar expansion nor anisotropic {stress}, and the temperature is related to the fluid acceleration, respectively.
 
Once the equilibrium state is defined, we now decompose each physical quantity into its equilibrium term plus its dissipative contribution. So far, we have presented a generic relativistic fluid, following a tensor decomposition into three scalars $n$, $\varepsilon$, and $\mathcal{P}$, two vectors $u_\mu$ and $\mathcal{Q}_{\mu}$, and a traceless tensor $\mathcal{T}_{\mu \nu}$. 
In the Eckart frame, the number density $n$, and the energy density $\varepsilon$, correspond to the thermodynamic quantities defined in equilibrium.
In principle, the rest of generic quantities have both {equilibrium and dissipative contributions. 
However, as the fluid in equilibrium is isotropic, $\mathcal{Q}_{\mu} = \mathcal{T}_{\mu \nu} = 0$, then $\mathcal{Q}_\mu$ and $\mathcal{T}_{\mu \nu}$ must be dissipative \cite{Israel:1979wp}. 
Therefore, the rest of generic quantities from the fluid in Eq. \eqref{eq:current_and_stress} match to
\begin{subequations}
    \label{eq:npf_decomposition}
    \begin{align}
    \mathcal{P} &= p + \Pi, \label{eq:def_pressure}\\
    \mathcal{Q}_\mu &= q_\mu, \\
    \mathcal{T}_{\mu \nu} &= \varPi_{\mu \nu}.
    \end{align}
\end{subequations}
Above, the isotropic pressure of the fluid $\mathcal{P}$ is split into the equilibrium pressure $p$, as in Eq. \eqref{eq:fund_law}, and the viscous bulk pressure $\Pi$. The vector $\mathcal{Q}_\mu$ corresponds to the heat flux $q_{\mu}$, while $\mathcal{T}_{\mu \nu}$ coincides with the anisotropic stress $\varPi_{\mu \nu}$. The scalar quantities $\{n, \varepsilon, p\}$ describe the fluid in equilibrium, and are related by the fundamental law of thermodynamics and the EoS. The quantities $\{\Pi, q_{\mu}, \varPi_{\mu \nu}\}$ are generically referred to as the dissipative fluxes, and are governed by irreversible thermodynamics.

\subsection{Extended irreversible thermodynamics} \label{sec:irre_thermo}
Irreversible processes must obey the second law of thermodynamics, involving an entropy increase. As for the conservation of the current and stress tensor, a conservation relation of the entropy flux $\mathbf{S}$ describes this law,
\begin{equation}
    \tilde \nabla^{\mu} S_{\mu} \geq 0. \label{eq:entropy}
\end{equation}
If the fluid is in equilibrium, the entropy flux is conserved, $\tilde \nabla^{\mu} S_{\mu} =0$.  
Thus, the dissipative terms increase the entropy of the system.  The entropy flux $\mathbf{S}$ adopts the following form,
\begin{equation}
    S^{\mu} = snu^{\mu} + \frac{R^{\mu}}{T}, \label{eq:entropy_flux}
\end{equation}
where the vector $R^\mu$ contains only the dissipative fluxes. Thus, the first term in {Eq. \eqref{eq:entropy_flux}} corresponds to the equilibrium quantity of the entropy flux. 
The local entropy of the fluid is the entropy measured by an observer comoving with the fluid. Then, projecting the entropy flux in Eq. \eqref{eq:entropy_flux} to the velocity $u^{\mu}$ yields the local entropy,
\begin{equation}
    -u_{\mu}S^{\mu} = sn - R^{\mu}u_{\mu}.
\end{equation}
In equilibrium, the dissipative flux vanishes, $R^{\mu}=0$, and the local entropy measured corresponds to the entropy density in equilibrium, $sn$, with the quantities used in the fundamental law in Eq. \eqref{eq:fund_law}. 

The definition of the dissipative flux $R^{\mu}$ depends on the choice of theory for irreversible thermodynamics.  
The MIS phenomenological definition \cite{Muller:1967zza,Israel:1979wp} is,
\begin{eqnarray}
    R^{\mu} &\equiv& q^{\mu} - \frac{1}{2} \left( \beta_1 \Pi^2 + \beta_2 q_\alpha q^\alpha + \beta_3 \varPi_{\alpha \beta} \varPi^{\alpha \beta} \right) u^{\mu} \nonumber \\
    &+& \alpha_1 \Pi q^{\mu} + \alpha_2 \varPi^\mu_\beta q^\beta, \label{eq:def_R}
\end{eqnarray}
where $\beta_1$, $\beta_2$, and $\beta_3$ are the thermodynamic coefficients that characterize dissipative contributions to the entropy flux in Eq. \eqref{eq:entropy_flux}. The coefficients $\alpha_1$ and $\alpha_2$ describe the coupling between fluxes. This definition of the dissipative flux $R^{\mu}$ is known as extended irreversible thermodynamics \cite{Maartens:1996vi}.

The positive increase of the entropy in Eq.\eqref{eq:entropy} imposes an evolution condition on the dissipative fluxes. The phenomenological equations that result from the application of this condition to the MIS definition in Eq. \eqref{eq:def_R}, are the MIS equations. 
By neglecting the coupling between fluxes, $\alpha_{1}=\alpha_{2}=0$, the vorticity of the system, $\omega_{\mu \nu} = 0$ (vanishing in spherical symmetry), and the gradient of the coefficients, these equations adopt the Maxwell-Cattaneo form \cite{Maartens:1996vi}
\begin{subequations}
    \label{eq:Mis_equations}
\begin{align}
    \tau_\zeta u^\lambda \tilde \nabla_\lambda \Pi + \Pi &= - \zeta \Theta, \label{eq:MIS_bulk}\\
    \tau_\kappa \Delta^{\alpha}_\mu u^\gamma \tilde \nabla_\gamma q_\alpha + q_\mu &= - \kappa T \left( \mathcal{D}_\mu \ln T + a_\mu \right), \label{eq:MIS_heat}\\
    \tau_\lambda \Delta^\alpha_\mu \Delta^\beta_\nu u^\gamma \tilde \nabla_\gamma \varPi_{\alpha \beta} + \varPi_{\mu \nu}&= - 2 \lambda \sigma_{\mu \nu}. \label{eq:MIS_shear}
\end{align}
\end{subequations}
The coefficients $\zeta$, $\kappa$ and $\lambda$ are the bulk viscosity, thermal conductivity, and shear viscosity, respectively, which correspond to the transport coefficients of the fluid.
Their particular values depend on the microscopic behavior of the fluid and the particles degrees of freedom used to describe matter i.e. hadrons or quarks. The coefficients $\tau_\zeta$, $\tau_\kappa$ and $\tau_\lambda$ are the relaxation times, which are related to the thermodynamic coefficients as  
\begin{equation} 
    \tau_\zeta \equiv \zeta \beta_1, \quad \tau_\kappa \equiv \kappa T \beta_2, \quad \tau_\lambda \equiv 2 \lambda \beta_3.
\end{equation}

The MIS equations describe the evolution of the dissipative fluxes. For a perfect fluid, the {transport} coefficients $\{\zeta, \kappa, \lambda\}$ {and their correspondent relaxation times $\{\tau_\zeta, \tau_\kappa, \tau_\lambda\}$} on Eqs. \eqref{eq:Mis_equations} are zero by definition, then, {there are no dissipative contributions, {$\Pi = q_\mu = \varPi_{\mu \nu} =0$}}.
Instead, for a non-perfect fluid the r.h.s. of the Eqs. \eqref{eq:Mis_equations} is only zero when the expansion scalar $\Theta$, the anisotropic tensor $\sigma_{\mu \nu}$, or $\left( \mathcal{D}_{\mu}\ln T + a_{\mu}\right)$ vanish. These conditions correspond to the equilibrium state in Eqs. \eqref{eq:equilibrium_conds}. In such state, any perturbation of the dissipative fluxes decays with a characteristic relaxation time.

{In summary, the conservation Eqs. \eqref{eq:conservation_laws} and the MIS Eqs. \eqref{eq:Mis_equations} describe the dynamics of a non-perfect fluid, once an EoS, and subsequently, the dissipative coefficients $\zeta$, $\kappa$, and $\lambda$, and their corresponding relaxation times are set.}

\section{Gauge-invariant perturbations in  spherical spacetime} \label{sec:GS_formalism}
We introduce the Gerlach-Sengupta formalism \cite{Gerlach:1979rw,Gerlach:1980tx} of gauge-invariant perturbations in a spherical background that contains an arbitrary type of matter. This approach applies the symmetry of the background to do a (2+2) split of spacetime. 
As a consequence, the dynamics of the perturbations depends on a decomposed 2-dimensional spacetime, which can be associated to the usual time and radial coordinates. In addition, using this formalism provides the benefit of being completely covariant and gauge-invariant.

\subsection{Spherical background}
We work on a spherically symmetric spacetime $\mathcal{M}$($\mathbf{g},~\tilde \nabla$). The symmetry of the system enables a decomposition of the manifold into a product of submanifolds $\mathcal{M} = \mathcal{M}_s \times S^2$. The time-radial submanifold $\mathcal{M}_s$ is described by a metric tensor $\mathbf{\Sigma}$, with its coordinates written with lowercase Latin letters as $x^a$ ($a=0,1$, in $\mu$ coordinates). 
The manifold $S^2$ corresponds to the unit 2-sphere with $x^a= x_0^a$ constant, described by the metric tensor $\mathbf{\Omega}$, and its coordinates are indicated by uppercase Latin letters $x^A$ ($A=2,3$, in $\mu$ coordinates).

Following the spherical background decomposition, we express  the background metric tensor $\bar{\mathbf{g}}$ as 
\begin{eqnarray}
	\bar{\mathbf{g}} &=& \bar g_{\mu \nu} \der x^\mu \der x^\nu \nonumber\\
    &=& \Sigma_{ab} \der x^a \der x^b + r^2(x^a) \Omega_{AB} \der x^A \der x^B, \label{eq:sph_metric_bg}
\end{eqnarray} 
with $r(x^a)$ a function on the time-radial submanifold $\mathcal{M}_s$. The metric of the 2-sphere is represented in standard angular coordinates  $\Omega_{AB} = \text{diag}(1, \sin^2\theta)$. The coordinates for the time-radial submanifold can be chosen as the Schwarzschild coordinates ($t,r$) with $t$ the coordinate time and $r$ the radius of the spherical system. Other charts may also be used for an explicit coordinate expression of $\mathbf{\Sigma}$, like the Kruskal-Szekeres or null coordinates \cite{Martel:2005ir}. However, the GS formalism allows working with covariant quantities to simplify the equations for the linear perturbations of any field on $\mathcal{M}$.

The full metric and each submetric have a compatible covariant derivative, which satisfy \cite{Gundlach:1999bt} 
\begin{equation} 
    \tilde \nabla_{\mu} g_{\nu \rho} = 0, \qquad    \nabla_{a} \Sigma_{bc} = 0, \qquad   D_A \Omega_{BC} = 0. 
\end{equation} 
The complete covariant derivative $\tilde \nabla$ is equivalent to $\nabla$ derivative for quantities only dependent on $x^a$ coordinates. In addition, we will use $(;)$ to express the $\mathcal{M}_s$ covariant derivative $\nabla$, as denoted above. 

The gradient of the function $r(x^a)$ defines a vector
\begin{equation}
	r^a \equiv \nabla^a r,
\end{equation} 
that describes the normal direction to the surfaces of the 2-sphere with constant $x^a$. 
We raise and lower its indices with the metric $\Sigma_{ab}$. The scalar product $r_a r^a$ is a covariant quantity with the form
\begin{equation}
	r^a r_a = 1 - \frac{2m(r)}{r}, \label{eq:def_f}
\end{equation}
from the $g_{rr}$ metric component in Schwarzschild coordinates. The mass $m$ in Eq. \eqref{eq:def_f} is equivalent to the Misner-Sharp mass
\cite{Gundlach:1999bt}, that is, the active gravitational energy of the spherical system. 

\noindent The Ricci tensor for the full spherical metric $\mathbf{g}$ is
\begin{subequations}
    \begin{align}
        ^{(4)} R_{ab} &= \mathcal{R}_{ab} - \frac{2}{r} r_{;ab},\\
        ^{(4)} R_{AB} &= \left(1 - r \square r - r_a r^a \right) \Omega_{AB},
    \end{align}
\end{subequations}
where $\square \equiv \Sigma^{ab}\nabla_{a} \nabla_{b}$ is the d'Alembert operator for time-radial coordinates, $\mathcal{R}_{ab}$ the curvature of the $\mathcal{M}_s$ manifold, and the curvature scalar is $\mathcal{R}\equiv \Sigma^{ab} \mathcal{R}_{ab}$. The Ricci scalar is,
\begin{equation}
    ^{(4)}R = \mathcal{R} + \mathfrak{R},
\end{equation}
with $\mathfrak{R} \equiv \frac{2}{r^2} \left(1 - 2 r \square r - r_a r^a\right)$. Then, the Einstein tensor components are,
\begin{subequations}
\begin{eqnarray}
    ^{(4)}G_{ab} &=& - \frac{2}{r} r_{;ab} - \frac{\mathfrak{R}}{2} \Sigma_{ab}, \label{eq:einsteinBG1} \\
    ^{(4)}G_{AB} &=& \left( r \square r - \frac{1}{2}r^2 \mathcal{R}\right) \Omega_{AB} . \label{eq:einsteinBG2}
\end{eqnarray}\label{eq:einsteinBG}
\end{subequations}

The Einstein tensor is sourced by spherically symmetric distribution of matter described by a stress-energy tensor as
\begin{equation}
	\bar{\mathbf{T}} = t_{ab} \der x^a \der x^b + \bar Q r^2 \Omega_{AB} \der x^A \der x^B, \label{eq:sph_stress_tensor_bg}
\end{equation}
with $t_{ab}(x^a)$ the time-radial part of the stress tensor and $\bar Q(x^a)$ the elements on the diagonal for the 2-sphere. The dependence on the complete set of $x^a$ coordinates allows the description beyond the equilibrium condition for background matter. 

From the Einstein equations, we obtain a set of covariant quantities using the generic stress tensor under spherical symmetry conditions in Eq. \eqref{eq:sph_stress_tensor_bg},
\begin{subequations}
    \label{eq:R_relations}
    \begin{align}
	   	\frac{r_{;ab}}{r}&= \frac{m}{r^3} \Sigma_{ab} - 4 \pi \left( t_{ab} - t \Sigma_{ab} \right), \\
	    \frac{\square r}{r} &=  \frac{2m}{r^3} + 4 \pi t,\\
        \mathcal{R} &= \frac{4m}{r^3} + 8 \pi ( t - 2 \bar Q), \\
        \mathfrak{R} &= - \frac{4m}{r^3} - 16 \pi t,
    \end{align} 
\end{subequations}
with $t\equiv t_{ab}\Sigma^{ab}$ the contraction of $t_{ab}$. The EFE are implicit in these relations, and together with the relation in Eq. \eqref{eq:def_f} describe the spherical background sourced by $\mathbf{T}$. These covariant quantities will be used to simplify the perturbed Einstein equations.

\subsection{Non-spherical perturbations}
We describe the perturbations as small displacements from the spherical configuration. Then, any tensor field on $\mathcal{M}$ is perturbed as a background quantity $\mathbf{X}$, only dependent on the time-radial coordinates, plus non-spherical perturbations $\delta \mathbf{X}$ as,
\begin{equation}
    \mathbf{X}(x^a, x^A) \to \mathbf{X}(x^a) + \delta \mathbf{X} (x^a, x^A).
\end{equation}
The spherical symmetry of the background allows a natural decomposition of any perturbation $\delta \mathbf{X}$, as the product of a scalar, vector or tensor  field on $\mathcal{M}_s$, times a scalar, vector or tensor spherical harmonic spanning $S^2$ \cite{Gerlach:1979rw}. We now detail each type of the spherical harmonics that we use in our derivations.

The scalar spherical harmonics, $Y_{lm}$, satisfy
\begin{equation}
	\Omega^{AB}\left( D_A D_B + \frac{l(l+1)}{2} \Omega_{AB} \right)Y_{lm} = 0. \label{eq:scalar_harmonics}
\end{equation}     

The vector spherical harmonics are defined as a set of two orthogonal elements  $\big[V^\mathcal{E}\big]$ and $\big[V^\mathcal{B} \big]$, that transform under parity as polar { or electric, $\mathcal{E}$,  and axial or magnetic, $\mathcal{B}$}, respectively. Their components are {defined for $l\geq 1$ as}
\begin{subequations}
	\begin{eqnarray}
    	\big[V^\mathcal{E}\big]_A &=& D_A Y_{lm}, \\
		\big[V^\mathcal{B}\big]_A &=& -\epsilon_{A}^B D_B Y_{lm},
	\end{eqnarray}
\end{subequations}
{with} $\epsilon_{AB}$ the Levi-Civita antisymmetric tensor of $S^2$. They satisfy the following relations concerning the angular covariant derivative,
\begin{subequations}
   \begin{eqnarray} 
    \Omega^{AB} D_B \big[V^\mathcal{E}\big]_A &=& - l(l+1) Y_{lm}, \\
    \Omega^{AB} D_B \big[V^\mathcal{B}\big]_A &=& 0.
    \end{eqnarray}
\end{subequations}   

Finally, the tensor spherical harmonics are defined as a contribution from a trace component $\big[ T^\mathcal{T}_{lm}\big]_{AB}$ and two traceless tensors $\big[ T^\mathcal{E}_{lm}\big]_{AB}$ and $\big[T^B_{lm}\big]_{AB}$. The first two transforms as polar, and the last one as axial. Their components are,
\begin{subequations}    
    \label{eq:tensor_harmonics}
    \begin{eqnarray}
        \big[ T^\mathcal{T}_{lm}\big]_{AB} &=& \Omega_{AB} Y_{lm}, \\
        \big[ T^\mathcal{E}_{lm}\big]_{AB} &=& \left( D_{A}D_B + \frac{1}{2}l(l+1)\Omega_{AB} \right)Y_{lm}, \\
        \big[ T^\mathcal{B}_{lm}\big]_{AB} &=& -\epsilon^{C}_{\,(A } D_{B)} D_C Y_{lm}. 
    \end{eqnarray}
\end{subequations}
The trace component $\big[T^\mathcal{T}_{lm} \big]$ is defined for $l\geq 0$, however, the traceless components $\big[ T^\mathcal{E}_{lm}\big] $ and $\big[ T^\mathcal{B}_{lm}\big]$ are only defined for the $l\geq 2$ case. This is the tensor decomposition used by Zerilli \cite{Zerilli:1970wzz} for black hole perturbations, and it is different from the choice by Regge and Wheeler in \cite{Regge:1957td} or Thorne and Campolattaro in \cite{thorne1967}. The use of the aforementioned decomposition has the advantage of constituting a set of orthogonal tensors, while $\big[ T^\mathcal{E}_{lm}\big]$ and $\big[ T^\mathcal{B}_{lm}\big]$ being traceless due to the properties of the scalar harmonics for the $\mathcal{E}$ component, and the antisymmetry of the $\mathcal{B}$ component in Eq. \eqref{eq:tensor_harmonics}. 

This decomposition in spherical harmonics  splits any perturbation over the spherical background into a time-radial and angular dependence. Then, the dynamics of the perturbation are on the time-radial contribution, simplifying the problem. 

We set the perturbations of the spacetime sourced by the perturbations of arbitrary matter. The general perturbation fields for the metric in Eq. \eqref{eq:sph_metric_bg}, and the matter stress-energy tensor in Eq. \eqref{eq:sph_stress_tensor_bg} are written as,
\begin{subequations}
    \begin{align}
        \delta \mathbf{g} = \delta g_{\mu \nu} \der x^\mu \der x^\nu
            =& \delta \Sigma_{ab} \der x^a \der x^b + \delta \gamma_{AB} \der x^A \der x^B \nonumber \\ 
            &+ \eta_{aB} \der x^a \der x^B, \\
        \delta \mathbf{T} = \delta T_{\mu \nu} \der x^\mu \der x^\nu
            =& \delta T_{ab} \der x^a \der x^b + \delta T_{AB} \der x^A \der x^B \nonumber \\ &+ \delta T_{aB} \der x^a \der x^B.
    \end{align}
\end{subequations}
The tensors $\delta \Sigma_{ab}(x^a, x^A)$, $\delta \gamma_{AB}(x^a, x^A)$, and $\eta_{aB}(x^a, x^A)$ provide the decomposition of the linear perturbations of the metric {$\mathbf{\delta g}$}. 
{Equivalently, }the tensors $\delta T_{ab}(x^a, x^A)$, $\delta T_{AB}(x^a, x^A)$, and $\delta T_{aB}(x^a, x^A)$ are the {decomposed} linear perturbations of the stress-energy tensor of matter {$\mathbf{\delta T}$}. Then, these general linear perturbations of $\delta g_{\mu \nu}$ are decomposed with the previously defined set of spherical harmonics as
\begin{subequations}
    \label{eq:pert_metric}
    \begin{align}
        \delta \Sigma_{ab} =& \sum_{l,m}h^{lm}_{ab} Y_{lm},  \label{eq:pert_metric_ab}\\
        \eta_{aA} =& \sum_{l,m} \big(  h_a^{lm} \big[V^\mathcal{E}_{lm}\big]_A + k_a^{lm} \big[ V^\mathcal{B}_{lm}\big]_A \big),  \label{eq:pert_metric_aB}\\
        \delta \gamma_{AB} =& r^2 \sum_{l,m} \big(   H^{lm} \big[T^\mathcal{T}_{lm}\big]_{AB} + G^{lm} \big[T^\mathcal{E}_{lm}\big]_{AB} \nonumber \\
    & \qquad + K^{lm} \big[T^\mathcal{B}_{lm}\big]_{AB}\big) .\label{eq:pert_metric_AB}
    \end{align} 
\end{subequations}
Accordingly, the stress-energy perturbations $\delta T_{\mu \nu}$ are,
\begin{subequations}
    \label{eq:stress_tensor_decomposition} 
    \begin{align}
        \delta T_{ab} =& \sum_{l,m}Q^{lm}_{ab} Y_{lm}, \label{eq:sress_tensor_ab}\\
        \delta T_{aA} =& \sum_{l,m} \left( Q_a^{lm} \big[V^\mathcal{E}_{lm}\big]_A +  S_a^{lm} \big[V^\mathcal{B}_{lm}\big]_A \right), \label{eq:sress_tensor_aB} \\
        \delta T_{AB} =& r^2 \sum_{l,m}\big(  Q^{lm} \big[T^\mathcal{T}_{lm}\big]_{AB} + P^{lm} \big[T^\mathcal{E}_{lm}\big]_{AB} \nonumber \\
        & \qquad + L^{lm} \big[T^\mathcal{B}_{lm}\big]_{AB} \big). \label{eq:sress_tensor_AB}
    \end{align}
\end{subequations}
From now on, the summation over angular momentum values ($l,m$) is implicit, unless otherwise stated.

The tensors $\{h_{ab}, h_a, H, G\}$ and $\{Q_{ab}, Q_a, Q, P\}$ belong to the polar sector, while, $\{k_a, K\}$ and $\{S_a, L\}$ to the axial one. These tensors are only dependent on the $x^a$ coordinates, as the $x^A$ dependence is in the spherical harmonics.
Due to the linearity of the perturbations, both sectors evolve independently to linear order.

\subsection{Gauge-invariant perturbations}
We aim to build gauge-invariant perturbation fields for the metric and stress-energy tensor to linear order. The perturbation of a given field $X$ transforms under an infinitesimal coordinate displacement $x^\mu \to x^\mu + \xi^\mu$ with the Lie derivative $\mathcal{L}_\xi \bar X$ of the background  value $\bar X$ as,
\begin{equation}
    \delta X \to \delta X + \mathcal{L}_\xi \bar X.
\end{equation}
Thus, the perturbation $\delta X$ is gauge-invariant to linear order if the Lie derivative of the background field vanishes i.e.,  $\mathcal{L}_\xi \bar{X} = 0$. For the spherically symmetric case, {we use gauge-invariant perturbations for}  $l \geq 2$, while for $l=0,1$ the perturbations are not gauge-invariance, and they need to be analyzed separately.

Let the coordinate displacement 4-vector be $\xi_{\mu}$. The Lie derivative of a generic rank-2 tensor field $X_{\mu \nu}$, e.g., the metric tensor $g_{\mu \nu}$ or the stress-energy tensor $T_{\mu \nu}$, is
\begin{equation}
    \mathcal{L}_{\xi} X_{\mu \nu} = \xi^\alpha \tilde \nabla_\alpha X_{\mu \nu} + X_{\mu \alpha} \tilde \nabla_\nu \xi^\alpha + X_{\nu \alpha} \tilde \nabla_\mu \xi^\alpha,
\end{equation}
where the first term vanishes {in the case the tensor $X_{\mu \nu}$} corresponds to the metric, $X_{\mu \nu} =g_{\mu \nu}$.
In order to study how each component of the perturbation transforms, we decompose the 4-vector $\xi_\mu$ as 
\begin{equation*}
    \xi_{\mu} = (\xi_a, \xi_A),
\end{equation*}
with, 
\begin{subequations}
    \begin{align}
        \xi_a &= \xi^{lm}_a Y_{lm}, \\ 
        \xi_A &= r^2 \left(\xi^{(\mathcal{E})}_{lm} \big[V_{lm}^\mathcal{E}\big]_A + \xi^{(\mathcal{B})}_{lm} \big[V_{lm}^\mathcal{B}\big]_A \right),
    \end{align}    
\end{subequations}
where $\xi^{(\mathcal{E})}$ and $\xi^{(\mathcal{B})}$ correspond to the terms in the polar and axial sector, respectively.

We apply the transformation to the components of the metric perturbation in Eq.~\eqref{eq:pert_metric} and stress-energy perturbation in Eq.~\eqref{eq:stress_tensor_decomposition}, as in\cite{Gerlach:1979rw}. In our notation, the transformations in the polar sector are,
\begin{align}
    (\delta \mathbf{g}) : &
    \begin{cases}
        h_{ab} &\to h_{ab} - 2 \xi_{(a;b)}, \\
        h_a &\to h_a - \xi_a - r^2  \xi^{(\mathcal{E})}_{\ ,a}, \\
        H &\to H + l(l+1) \xi^{(\mathcal{E})} - 2 \frac{r^{a}}{r} \xi_{a}, \\
        G &\to G - 2 \xi^{(\mathcal{E})},    
    \end{cases}\\
    (\delta \mathbf T) : &
    \begin{cases}
        Q_{ab} &\to  Q_{ab} - \xi^c t_{ab;c} - 2 \xi^c_{(;a} t_{b)c}, \\
         Q_a &\to  Q_a - t_{ab} \xi^b - r^2 \bar Q \xi^{(\mathcal{E})}_{\ ,a}, \\
         Q &\to  Q + l(l+1) \bar Q \xi^{(\mathcal{E})} - \frac{\xi^a}{r^2} \left( r^2\bar{Q} \right)_{,a} , \\
        P &\to  P - 2\bar Q \xi^{(\mathcal{E})},
    \end{cases}
\end{align}
where  we have ignored the $l,m$ labels for each quantity.
With these transformations, we take the gauge-invariant metric and matter perturbations as,
\begin{align}
    \label{eq:gi_polar_grav}(\delta \mathbf{g}): &
    \begin{cases}
        \tilde h_{ab} &= h_{ab} - 2\epsilon_{(a;b)} , \\
        \tilde H &= H + \frac{l(l+1)}{2} G  - 2 \frac{r^a}{r} \epsilon_a ,
    \end{cases} \\
    \label{eq:gi_polar_matter} (\delta \mathbf{T}): &
    \begin{cases}
        \tilde Q_{ab} &= Q_{ab} - \left( t_{ab;c} \epsilon^c + 2 \epsilon^c_{(;a} t_{b)c} \right), \\
        \tilde Q_a &= Q_a - t_{ac}\epsilon^c - \frac{r^2}{2} \bar Q G_{,a}, \\
        \tilde Q & = Q + \frac{l(l+1)}{2} \bar Q G - \frac{1}{r^2} \left( r^2 \bar{Q}\right)_{,c} \epsilon^c , \\
        \tilde P &= P - \bar Q G,
    \end{cases}
\end{align}
with $    \epsilon_a \equiv h_a - \frac{r^2}{2} G_{,a} \label{eq:epsilon_pert}$.

In the polar sector, we have four degrees of freedom in the gauge-invariant metric perturbations given by a symmetric tensor $\tilde h_{ab}$ and a scalar $\tilde H$. For matter perturbations, we have seven degrees of freedom in the gauge-invariant stress-energy tensor due to the symmetric tensor $\tilde Q_{ab}$, the vector $\tilde Q_a$ and two scalars $\tilde Q$ and $\tilde P$.

For the axial sector, the transformations are
\begin{align}
    (\delta \mathbf{g}) : & 
    \begin{cases}
        k_a &\to k_a - r^2 \xi^{(\mathcal{B})}_{\ ,a}, \\
        K &\to K - 2 \xi^{(\mathcal{B})},            
    \end{cases} \\
    (\delta \mathbf{T}) : &
    \begin{cases}
        S_a \to S_a - r^2 \bar Q \xi^{(\mathcal{B})}_{\ ,a}, \\
        L \to L - 2 \bar Q \xi^{(\mathcal{B})},
    \end{cases}
\end{align}
where we have dropped the $l$, $m$ labels as well. Similarly, we build a set of gauge-invariant quantities for axial metric and matter perturbations as
\begin{align}
    \label{eq:gi_axial_grav}(\delta \mathbf{g}) : & 
        \quad \tilde k_a = k_a - \frac{1}{2} r^2 K_{,a}, \\
    \label{eq:gi_axial_matter}(\delta \mathbf{T}) : &
    \begin{cases}
        \tilde S_a = S_a - \bar{Q} k_a, \\
        \tilde L\phantom{_a} = L - \bar{Q} K.
    \end{cases}
\end{align}
In the axial sector, we have two degrees of freedom in the gauge-invariant metric perturbations from a vector $\tilde k_a$. For matter perturbations, we have three degrees of freedom in the gauge-invariant stress tensor given by a vector $\tilde S_a$ and a scalar $\tilde L$.

The gravitational pertubations built in Eq. \eqref{eq:gi_polar_grav} and Eq. \eqref{eq:gi_axial_grav} are only gauge-invariant for the $l\geq 2$ case, as for $l=0,1$, some perturbations are not defined. In these cases, the gauge degrees of freedom are not saturated, and additional gauge fixing can be imposed. The polar $l=0$ perturbations can be understood as redefinition of the background metric, while the polar $l=1$ perturbations corrspond to a coordinate displacement and can be removed by a gauge transformation. The axial $l=1$ perturbations are determined by the matter perturbations, having no degrees of freedom \cite{Zerilli:1970wzz,Martel:2005ir}. Thus, we work with the expression for any $l$, focusing on the gauge-invariant radiative cases $l\geq 2$, and then address the special case of the lower-order perturbations in Sections \ref{sec:polar_low} and \ref{sec:axial_low}.

The usual gauge choice to study perturbations on a spherical system is the Regge-Wheeler (RW) gauge \cite{Regge:1957td}, which sets $h_a^{\text{RW}} = G^{\text{RW}} = K^{\text{RW}}= 0$. This choice fixes the metric perturbations to the gauge-invariant quantities, $\tilde h_{ab} = h_{ab}^{\text{RW}}$, $ \tilde k_a = k_a^{\text{RW}}$, $\tilde H = H^{\text{RW}}$. Therefore, working in the RW gauge is equivalent to doing so with gauge-invariant metric perturbations. For {polar} matter perturbations, we have $\tilde Q = Q_{ab}^{\text{RW}}$, $\tilde Q_a = Q_{a}^{\text{RW}}$, $\tilde Q = Q^{\text{RW}}$, and $\tilde P = P^{\text{RW}}$, thus, the RW gauge is equivalent to work with gauge-invariant matter perturbations.
{However, this is not the case for the axial matter perturbations, as }
$\tilde S_a = S_a^{\text{RW}} - \bar Q k_a^{\text{RW}}$ and $\tilde L = L^{\text{RW}}$. The vector $S_a$ has a contribution from the metric $k_a^{\text{RW}}$, so, in order to work with a set of gauge-invariant perturbations, we define $L_a \equiv \tilde S_a = S_a^{\text{RW}} - \bar Q k_a^{\text{RW}}$ as the gauge-invariant vector for axial matter perturbations.

From now on, we use the RW gauge, as it provides gauge-invariant metric and axial matter perturbations. {Instead,} for {polar} matter perturbations, we use $L_a$ and $\tilde L$. In addition, we drop the tilde symbols from the gauge-invariant perturbations to lighten the notation. 

\subsection{Perturbed Einstein equations}
We now focus on the Einstein tensor perturbed to linear order and decompose into a background part corresponding to Eq.~\eqref{eq:einsteinBG}, and a perturbation $\delta G_{\mu \nu}$. In GR, the perturbed Einstein tensor for a non-vanishing background curvature reads
\begin{equation}
    \delta G_{\mu \nu} =  \delta R_{\mu \nu} - \frac{1}{2} g_{\mu \nu} \delta R - \frac{1}{2} \delta g_{\mu \nu} \bar R,
\end{equation}
with the perturbed Ricci tensor,
\begin{equation}
    \delta R_{\mu \nu} =  \tilde \nabla_\alpha \delta \Gamma^{\alpha}_{\mu \nu} - \tilde \nabla_\nu \delta \Gamma^{\alpha}_{\, \mu \alpha}.
    \label{eq:RicciTensor}
\end{equation}
and $\delta R \equiv g^{\mu \nu} \delta R_{\mu \nu} - \delta g^{\mu \nu} \bar R_{\mu \nu}$ the perturbed Ricci scalar. $\bar R_{\mu \nu}$ and $\bar R$ correspond to the background Ricci tensor and Ricci scalar, respectively. In addition, the unperturbed Christoffel symbols are 
\begin{equation}
    \Gamma^{\mu}_{\, \nu \sigma} \left(\mathbf{g}\right) = \frac{1}{2} g^{\mu \alpha} \left( \tilde \nabla_\nu g_{\sigma \alpha} + \tilde \nabla_\sigma g_{\nu \alpha} - \tilde \nabla_\alpha g_{\nu \sigma} \right).
\end{equation}
{Then, the perturbed Christoffel symbols result}
\begin{equation}
    \delta \Gamma^{\mu}_{\nu \sigma} = \tilde \nabla _{(\nu} h^{\mu}_{\sigma)} - \frac{1}{2} \tilde \nabla^\mu h_{\nu \sigma}  -  h^{\mu}_{\ \alpha} \bar \Gamma ^{\alpha}_{ \nu \sigma} ,
\end{equation}
with the covariant derivative compatible with the background metric. 

{Therefore, the perturbed Einstein equations, using the decomposition of the perturbed metric tensor in Eq. \eqref{eq:pert_metric}, and the perturbed stress tensor in Eq. \eqref{eq:stress_tensor_decomposition}, read as }
\begin{widetext}
\begin{subequations}
    \label{eq:einstein_pert_polar}
    \begin{align}
        \left(- \frac{\mathfrak{R}}{2}    + \frac{l(l+1)}{2r^2} \right)h_{ab}  + \frac{r^m}{r} \left( 2\nabla_{(a} h_{b)m}   -  \nabla_{m} h_{ab} + \Sigma_{ab}\left(  h_{,m} -2 \nabla^{n} h_{mn}\right) \right)& & \nonumber \\
        - \Sigma_{ab} \left( \frac{r^m r^n}{r^2}  h_{mn} -  8 \pi t_{mn} h^{mn} + \left( - \frac{\mathfrak{R} }{2}   + \frac{l(l+1)}{2r^2} \right) h \right)& &  \nonumber \\
        - \nabla_{ab}H - \frac{2}{r} r_{(a}H_{,b)} + \Sigma_{ab} \left( \square H + 3 \frac{r^c}{r} H_{,c} -  \frac{(l-1)(l+2)}{2r^2}H\right) &= 8 \pi Q_{ab} &(l \geq 0 ), \label{eq:einstein_pert_polar_a}\\
        \square h - \nabla_{mn} h^{mn} - \frac{2}{r}r^m \nabla_{n} h^n_{m} + \frac{r^a}{r}h_{,a} - \frac{l(l+1)}{2r^2} h + 8 \pi \left( t^{ab} - \frac{1}{2} \Sigma^{ab}( t + 2 \bar{Q}) \right) h_{ab}  \nonumber & & \label{eq:einstein_pert_polar_b}\\
        + 2\frac{r^m}{r} H_{,m} + \square H - \left( \frac{4m}{r^3} + 8 \pi ( t - 2 \bar Q) \right) H &= 16 \pi Q, &(l\geq 0), \\\nabla^m h_{ma} - h_{,a} + \frac{r_a}{r} h - H_{,a} &= 16 \pi Q_a & (l \geq 1), \label{eq:einstein_pert_polar_c}\\
        - \frac{1}{2} h &= 8 \pi P &(l \geq 2),\label{eq:einstein_pert_polar_d}
    \end{align}
\end{subequations}
where we have denoted the trace of $h_{ab}$ as $h \equiv \Sigma^{ab} h_{ab}$, and the following  relation among $h_{ab}$ on a 2-dimensional manifold with scalar curvature $\mathcal{R}$,
\begin{align}
    \nabla^c\left( h_{ab;c} - 2 h_{c(a;b)} \right) + h_{;ab} + g_{ab} \left( h_{mn}^{\ ; mn} - \square h\right) = \frac{\mathcal{R}}{2} ( h g_{ab} - 2 h_{ab} ),
\end{align}
and the covariant quantities in Eqs. \eqref{eq:R_relations} from the background Einstein equations. 
Equivalently, the axial sector results in two equations,
\begin{subequations}
    \label{eq:einstein_pert_axial}
    \begin{align}
        -\frac{1}{2} \nabla^c\Bigg[r^4  \nabla_c \left( \frac{k_a}{r^2}\right) - r^4 \nabla_a\left( \frac{k_c}{r^2}\right) \Bigg] + \frac{(l-1)(l+2)}{2} k_a  &=  8 \pi r^2 L_a &(l\geq 1), \\
        \nabla_m k^m &= 8 \pi r^2 L &(l \geq 2).
    \end{align}
\end{subequations}

These equations are {expressed using covariant and gauge-invariant perturbations on the 2-dimensional manifold $\mathcal{M}_s$}, {and show} the evolution of the metric perturbations from a given stress-energy tensor decomposed as in Eqs. \eqref{eq:stress_tensor_decomposition}. 
\end{widetext}

\section{non-perfect fluid in spherical spacetime} \label{sec:pert_fluid}
{In section \ref{sec:GS_formalism}, we reviewed the formalism to obtain the dynamics of metric perturbations {from} an arbitrary type of matter. In the current section, we particularize matter to be a relativistic {non-perfect fluid with dissipative effects}, as presented in section~\ref{sec:fluid}, in spherical spacetime. We consider the fluid described by an EoS of two general thermodynamic variables, e.g. $p(n, \varepsilon)$. In most astrophysical scenarios, two variables from  the set $\varepsilon$,  $n$,  $s$, are chosen as independent variables. As consequence, we present the calculations using explicitly those three variables. }
{However, only two of them are independent. }

We also assume that the dissipation coefficients such as the bulk $\zeta$ viscosity, shear viscosity $\lambda$, and the thermal conductivity $\kappa$, {as well as their correspondent relaxation times,} are parameters arising{ from the microphysics of the fluid\cite{Cerda2010}}.

\subsection{Background equations}
We follow the approach from Gundlach and Martín-García \cite{Gundlach:1999bt} describing  all the field equations only with scalar quantities.
In spherical symmetry, the time-like 4-velocity, $u_\mu$, has only a non-zero time-radial component, $u_{\mu}~ = \left(u_a ,0\right)$. This condition {implies an orthogonal space-like velocity { on } $\mathcal{M}_s$,} in the form $n_a =~- \epsilon_{ab} u^b$, with $\epsilon_{ab}$ the antisymmetric tensor. {We associate this $n_a$ direction to a radial-like direction given a proper set of coordinates.} {Then}, these two vectors satisfying
\begin{equation}
    \Sigma^{ab} u_a u_b = -1, \quad \Sigma^{ab} n_a n_b = 1, \quad \Sigma^{ab} u_a n_b = 0,
\end{equation}
describe an orthonormal basis for vectors in $\mathcal{M}_s$, and defines the frame derivatives,
\begin{equation}
    \dot f \equiv u^a \nabla_a f, \quad f' \equiv n^a \nabla_a f. 
\end{equation}

{These derivatives describe} the change of a given quantity $f$ measured by an observer following $u_a$ and $n_a$, respectively. {We understand the dot derivative ($\cdot$) as a comoving time derivative, and the prime derivative ($'$) as an orthogonal to time direction or radial-like derivative}. In addition, we also work with the following scalar contractions,
\begin{align}
    \mu &\equiv \nabla_a u^a,    &  \nu &\equiv \nabla_a n^a, \nonumber \\
    U &\equiv \frac{r_a}{r} u^a, &  W &\equiv \frac{r_a}{r} n^a. \label{eq:def_mu_nu}
\end{align}
The quantity $\mu$ ($\nu$) describes the expansion scalar following $u_a$ ($n_a$) in the $\mathcal{M}_s$ manifold, while $U$ ($W$) defines the scalar expansion from the scalar function $r$, following $u_a$ ($n_a$). The expansion scalar for the full 4-dimensional space is $\Theta=\tilde \nabla^\mu u_{\mu} = \mu + 2U$. {Furthermore}, the covariant derivatives of the vectors satisfy
\begin{align}
    \nabla_b u_a &= n_a \left( n_b \mu - u_b \nu\right), \nonumber\\
    \nabla_b n_a &= u_a \left( n_b \mu - u_b \nu\right). 
\end{align}

From the vector basis $\{u_a, n_a\}$, the following {symmetric} tensors form an orthogonal basis \cite{Gundlach:1999bt},
\begin{align}
    \Sigma_{ab} &= - u_a u_b + n_a n_b, \nonumber\\
    p_{ab} &= u_a u_b + n_a n_b, \nonumber \\
    q_{ab} &= n_a u_b + u_a n_b,
    \label{eq:tensorbasis}
\end{align}
with the tensor $\Sigma_{ab}$ corresponding to the metric of $\mathcal{M}_s$.

{With a vector and tensor basis characterized by the fluid velocity, we define the current vector and the stress tensor in the spherical system.}
The {density} current vector $J_\mu$ in the spherical background is
\begin{equation}
    J_{a} = n u_a, \quad J_A = 0. \label{eq:current_bg}
\end{equation}
We have chosen the Eckart frame ($\mathcal{J}_\mu = 0$), so the matter current flows along the worldline of $u_{\mu}$, and $n$ corresponds to the number density, as in Eq. \eqref{eq:npf_decomposition}. Imposing spherical symmetry, the stress-energy tensor for a non-perfect fluid in the spherical background reads
\begin{subequations}
    \label{eq:stressBG}
    \begin{align}
        \bar T_{ab} &= \varepsilon u_a u_b + \mathcal{P}_\parallel n_a n_b + \mathcal{Q} q_{ab}, \label{eq:stressBG1} \\
        \bar T_{AB} &= \mathcal{P}_\perp r^2 \Omega_{AB}, \label{eq:stressBG2}
    \end{align}        
\end{subequations}
where we have defined $\mathcal{P}_\parallel \equiv \mathcal{P} + \varPi_\parallel $ and $\mathcal{P}_\perp \equiv \mathcal{P} + \varPi_\perp$. We identify these quantities as the longitudinal pressure $\mathcal{P}_\parallel$, and the tangential pressure $\mathcal{P}_\perp$, {respect to the $n_a$ direction}. We recover the isotropic pressure with the combination $\mathcal{P} = \frac{1}{3}(\mathcal{P}_\parallel + 2 \mathcal{P}_\perp)$. 
The energy density is $\varepsilon$ and the {isotropic pressure, $\mathcal{P}$, corresponds to} the equilibrium pressure $p$ and the dissipative contribution, the viscous bulk pressure, $\Pi$, as in Eq. \eqref{eq:def_pressure}. 

{The heat flux $q_\mu$ in spherical symmetry has only a time-radial component $q_a$ that is parallel to the $n_a$ direction, $q_{a} = \mathcal{Q}n_a$}. Then, the scalar $\mathcal{Q}$ is the {magnitude of the } heat flux transferred in the $n_a$ direction. {The anisotropic stress $\varPi_{\mu \nu}$ in spherical symmetry is characterized by two quantities:} $\varPi_\parallel$ and $\varPi_\perp$ that account for the anisotropic viscous {stress }
contribution in the fluid. They fulfill the traceless condition of the anisotropic {stress} tensor,
\begin{equation}
    \varPi_\parallel + 2 \varPi_\perp = 0, \label{eq:bg_anis}
\end{equation}
so that we can set one of them, for example, $\varPi_\parallel$, to describe pressure anisotropies of the background fluid.

From the stress-energy tensor in Eqs. \eqref{eq:stressBG} and the current vector in Eq. \eqref{eq:current_bg}, we obtain the description of the fluid dynamics in the background from local conservation laws in Eq. \eqref{eq:conservation_laws}. In addition to the matter and energy conservation Eqs. (\ref{eq:cons_matter},~\ref{eq:cons_energy}), we project  the $\{u_a, n_a\}$ basis into the Euler equation \eqref{eq:cons_euler},
\begin{subequations}
    \label{eq:bg_cons}
    \begin{align}
        \dot{n} + \Theta n &= 0, \label{eq:currentcons} \\
        \dot{\varepsilon} + \mu \left( \varepsilon + \mathcal{P}_\parallel\right)  + 2U \left( \varepsilon + \mathcal{P}_\perp\right)& \nonumber \\
        + \mathcal{Q}' + 2 \left( W + \nu\right)\mathcal{Q} &= 0 ,\label{eq:conservation} \\
        \mathcal{P}_\parallel' + \nu \left( \varepsilon + \mathcal{P}_\parallel \right) + 2W \left( \mathcal{P}_\parallel - \mathcal{P}_\perp\right)& \nonumber \\
        + \dot{\mathcal{Q}} + 2 \left( U + \mu\right)\mathcal{Q}  &= 0. \label{eq:eulercons}
    \end{align}        
\end{subequations}
They correspond to the number, energy conservation and the Euler equation for the spherical system, {that describe the evolution of $n$, $\varepsilon$, and $\dot{\mathcal{Q}}$, respectively.}

{We also aim to work with the specific entropy $s$ as an independent variable.} Then, using the fundamental law of thermodynamics Eq. \eqref{eq:fund_law}, the change of entropy along the fluid worldlines is,
\begin{equation}
    \dot s =- \left( \Theta \Pi+ (\mu - U) \varPi_\parallel + \mathcal{Q}' + 2(\nu + W) \mathcal{Q}\right). \label{eq:entropy_cons_bg}
\end{equation}

These equations, supplemented with an EoS, completely determine the evolution of the system in the case of a perfect fluid ($\Pi=\mathcal{Q}= \varPi_\parallel = 0$). 
{However,} for a non-perfect fluid, we further need a set of constitutive equations for the dynamics of the dissipative quantities. 
We impose the spherical symmetry on the MIS equations in Eq. \eqref{eq:Mis_equations} so that they read
\begin{subequations}
    \label{eq:bg_MIS}
\begin{eqnarray}
    \tau_\zeta \dot \Pi + \Pi &=& - \zeta (\mu + 2 U)  , \\
    \tau_\kappa \dot{\mathcal{Q}} +  \mathcal{Q}&=& - \kappa T \left( \ln T' + \nu \right), \\
    \tau_\lambda \dot \varPi_\parallel + \varPi_\parallel &=& - \frac{4}{3}\lambda \left( \mu - U \right).
\end{eqnarray}
\end{subequations}

In the equilibrium state, the dissipative fluxes vanish, yielding,
\begin{subequations}
    \label{eq:equilibrium_bg}
    \begin{align}
        \mu = U &= 0, \label{eq:eq_mu}\\
        \nu + \ln T' &= 0.  \label{eq:Tolman}
    \end{align}        
\end{subequations}
Equation \eqref{eq:eq_mu} describes the background fluid as stationary. 
Equation \eqref{eq:Tolman} is the Ehrenfest–Tolman effect in equilibrium. These identities transform the conservation equations to,
\begin{align}
         \dot n = \dot \varepsilon = \dot s &= 0, \\ 
         p' + \nu \left( \varepsilon + p \right) &= 0. \label{eq:TOV}
\end{align}
The first equations are the local conservation of $n$, $\varepsilon$, and $s$ along the worldline of the fluid, while the second is the generalization of the Tolman-Oppenheimer-Volkoff (TOV) equation in the spherically symmetric formalism.

Setting the Einstein tensor in Eq. $\eqref{eq:einsteinBG}$ to $8 \pi \bar T_{ab}$ \eqref{eq:stressBG} we get the background EFE. We project these equations into the tensor basis in Eq. \eqref{eq:tensorbasis}, and  after some algebra, we obtain
\begin{subequations}
    \label{eq:BG_EFE}
    \begin{align}
        W' &= - W^2 + \mu U - 4\pi \mathcal{E} + \frac{m}{r^3}, \\
        \dot U &= -U^2 +\nu W - 4\pi \mathcal{P}_\parallel - \frac{m}{r^3}, \\
        \dot W &= U \left( \nu - W \right) + 4 \pi \mathcal{Q}, \\ 
        U' &= W \left( \mu - U \right) + 4 \pi \mathcal{Q}, \\
        \dot \mu &= - \mu^2 + \nu'  + \nu^2  \nonumber \\
        &+ 4\pi \left(- \mathcal{E} + \mathcal{P}_\parallel - 2 \mathcal{P}_{\perp}\right) + \frac{2m}{r^3}.
    \end{align}        
\end{subequations}

The Eqs. \eqref{eq:BG_EFE} describe the evolution and constrains of the spherical metric. Definition in Eq. \eqref{eq:def_f} leads to a relation between the mass $m$, U and W,
\begin{equation}
    \frac{2m}{r} = 1 + r^2\left( U^2 - W^2 \right).
\end{equation}

We have characterized the dynamics of the self-gravitating non-perfect fluid in spherical symmetry. The three equations \eqref{eq:currentcons}, \eqref{eq:conservation}, and \eqref{eq:entropy_cons_bg}, describe the evolution of the thermodynamic magnitudes, $n$, $\varepsilon$, and $s$, respectively. Due to the EoS and the first law of thermodynamics, only two of these quantities are independent, and one of these equations is redundant.  The three MIS equations \eqref{eq:bg_MIS} yield the evolution of the three dissipative magnitudes, $\Pi$, $\mathcal{Q}$, and $\varPi _{\parallel}$. Lastly, the Euler equation \eqref{eq:eulercons} is not an evolution equation, but a constraint on the longitudinal pressure $\mathcal{P}_{\parallel}$, describing the structure of the spherical system with a generalization of the TOV equation \eqref{eq:TOV}.
The EFE \eqref{eq:BG_EFE} determine the evolution of the spherical spacetime represented by the metric functions $\{U, W, \mu, \nu\}$.

\subsection{Non-spherical fluid perturbations}
We describe the perturbations as small displacements from the spherical configuration. The $\delta$-symbol indicates the perturbations up to first order. For the metric part, we use the gauge-invariant perturbations as described in section \ref{sec:GS_formalism}. Then, the perturbations $h_{ab}$ and $H$ describe the polar sector, while $k_a$ describes the axial sector. Using the tensor basis from Eq. \eqref{eq:tensorbasis} we characterize the polar perturbation $h_{ab}$ into three scalars $\eta$, $\phi$, $\psi$ as in \cite{Gundlach:1999bt}, 
\begin{equation}
    h_{ab} = \eta \Sigma_{ab} + \phi p_{ab} + \psi q_{ab}.
\end{equation}

For matter perturbations, we displace the generic stress tensor and current vector in Eq. \eqref{eq:current_and_stress} from the spherical background. 
The perturbations to first order of the density current and the stress-energy are
\begin{subequations}
    \begin{align}
    \delta J_{\mu} =& \delta n u_{\mu} + n \delta u_{\mu}, \\
    \delta T_{\mu \nu} =& 2\left( \varepsilon + \mathcal{P}\right) u_{(\mu}\delta u_{\nu)}  + \mathcal{P} \delta g_{\mu \nu} + 2 \mathcal{Q}_{(\mu} \delta u_{\nu)} \nonumber \\
    & + \left( \delta \varepsilon + \delta \mathcal{P} \right) u_{\mu} u_{\nu} + \delta \mathcal{P} g_{\mu \nu} + 2 \delta \mathcal{Q}_{(\mu} u_{\nu)} + \delta \mathcal{T}_{\mu \nu}.
\end{align}
\end{subequations}
Thus, the fluid perturbation from the current and stress tensor are represented by scalars $\delta n$, $\delta \varepsilon$, and $\delta \mathcal{P}$, two vectors $\delta u_{\mu}$ and $\delta \mathcal{Q}_{\mu}$, and a transverse traceless tensor $\delta \mathcal{T}_{\mu \nu}$. The perturbed scalars, {such as } $\delta n$, $\delta \varepsilon$, and $\delta \mathcal{P}$, define the perturbations from scalar and tensor traces quantities. We decompose any perturbed scalar quantity $\delta X$ ($l\geq 0$) as,
\begin{equation}
    \delta X = \delta X^{lm}(x^a) Y_{lm}(x^A)
\end{equation}
with the scalar harmonics $Y_{lm}$. We denote with $\delta X \equiv \delta X^{lm}(x^a)$ the field perturbation on $\mathcal{M}_s$, in order to lighten notation. The perturbed scalars have a direct match with the {perturbed} equilibrium thermodynamic quantities as the number density perturbation $\delta n$, energy density perturbation $\delta \varepsilon$, and the specific entropy perturbation $\delta s$. The perturbation of the general isotropic pressure $\delta \mathcal{P}$ includes the contribution of the equilibrium pressure perturbation $\delta p$ and the viscous pressure  perturbation $\delta \Pi$.

Since only two of the equilibrium quantities are independent and as it is usual practice, we use $\delta \varepsilon$ and $\delta s$ as the perturbations of the independent equilibrium variables of the system. Then, the perturbed equilibrium pressure depends on the EoS, and it is written as,
\begin{equation}
    \delta p = c_s^2 \delta \varepsilon + k_s \delta s, \label{eq:pert_pressure}
\end{equation}
with $c_s^2 \equiv \left( \tfrac{\partial p}{\partial \varepsilon} \right)_s$, and $k_s \equiv \left( \tfrac{\partial p}{\partial s} \right)_\varepsilon$, the sound speed of the fluid and the entropy coefficient, respectively. 

Regarding the vector quantities, the perturbation of the velocity is
\begin{equation}
    \delta u: \begin{cases}
        \delta u_a = \left(\frac{1}{2} \left( \eta - \phi \right) u_a + \left(\mathcal{V}_R - \frac{\psi}{2} \right) n_a \right) Y_{lm}, \\
        \delta u_A = \mathcal{V}_\mathcal{E} \big[V^\mathcal{E}]_A + \mathcal{V}_\mathcal{B}\big[V^\mathcal{B}]_A,
    \end{cases} \label{eq:vel_pert}
\end{equation}
where $\mathcal{V}_R$ ($l\geq 0$) corresponds to radial-like perturbations, $\mathcal{V}_\mathcal{E}$ to polar perturbations, and $\mathcal{V}_\mathcal{B}$ to axial perturbations of the fluid velocity.  We have imposed the normalization condition, $u_{\mu} u ^{\mu}= -1$, to linear order.

The {transversal condition of the fluid, $u^\mu \mathcal{Q}_{\mu} = 0$,} constraints the perturbation of the heat flux vector $\delta \mathcal{Q}_{\mu}$, which {fixes} the $u_a$ component of the flux perturbation. Then, the perturbation is
\begin{equation}
    \delta \mathcal{Q}:
    \begin{cases}
        \delta \mathcal{Q}_a = \left(\mathcal{Q} \left( \mathcal{V}_R + \frac{\psi}{2}\right) u_a + \delta \mathcal{Q}_R n_a \right) Y_{lm}, \\
        \delta \mathcal{Q}_A = \delta \mathcal{Q}_\mathcal{E} \big[V^\mathcal{E}]_A + \delta \mathcal{Q}_\mathcal{B} \big[V^\mathcal{B}]_A,    
    \end{cases} \label{eq:heat_flux_pert}
\end{equation}
where $\delta \mathcal{Q}_R$ ($l \geq 0$) corresponds to radial, $\delta \mathcal{Q}_\mathcal{E}$ ($l \geq 1$)  to polar, and $\delta \mathcal{Q}_\mathcal{B}$ ($l \geq 1$) to axial heat flow perturbations.

Finally, the  transversality condition, $u^\mu \mathcal{T}_{\mu \nu}=0$, restricts the perturbation of the anisotropic stress, 
\begin{equation}
    \delta \mathcal{T}:
    \begin{cases}
        \delta \mathcal{T}_{ab} = \left(\left(\delta \varPi_\parallel +  \left( \eta + \phi\right) \varPi_\parallel \right) n_a n_b \right. \nonumber \\ 
        \qquad \qquad +\left. \left( \mathcal{V}_R + \frac{\psi}{2}\right) \varPi_\parallel q_{ab} \right) Y_{lm},   \\
        \delta \mathcal{T}_{aA} = 
        \left(  \left( \mathcal{V}_\mathcal{E} - u_b h^b \right)\varPi_\perp u_a + \delta\varPi_{R\mathcal{E}} n_a \right)\big[V_{lm}^\mathcal{E}\big]_A \\
        \qquad \qquad + \left(  \left( \mathcal{V}_\mathcal{B} - u_b k^b \right)\varPi_\perp u_a + \delta\varPi_{R \mathcal{B}} n_a \right)\big[V_{lm}^\mathcal{B}\big]_A,  \\
        \delta \mathcal{T}_{AB} = r^2 \Big( \left( \delta \varPi_\perp + H\varPi_\perp\right)  \big[T_{lm}^\mathcal{T}\big]_{AB} \\
        \qquad \qquad + \delta\varPi_{\mathcal{E}} \big[T_{lm}^\mathcal{E}\big]_{AB} + \delta \varPi_{\mathcal{B}} \big[T_{lm}^\mathcal{B}\big]_{AB} \Big),
    \end{cases}
\end{equation}
and the traceless condition for $\delta \mathcal{T}_{\mu \nu}$,
\begin{equation}
    \delta \varPi_\parallel + 2 \delta \varPi_\perp = 0.
\end{equation}
The quantity $\delta \varPi_\parallel$ ($l \geq 0$) describes the anisotropic {longitudinal} pressure, $\delta \varPi_{R\mathcal{E}}$ ($l \geq 1$), and $\delta \varPi_\mathcal{E}$ ($l \geq 2$) characterize the radial-angular and angular anisotropic stress in the polar sector. The quantities  $\delta \varPi_{R\mathcal{B}}$ ($l \geq 1$), and $\delta\varPi_\mathcal{B}$ ($l \geq 2$) are their equivalent in the axial sector.

As for the background, the anisotropic perturbations $\delta \mathcal{Q}$ and $\delta \mathcal{T}$ are dissipative, and the dissipative scalars $\{\delta \mathcal{Q}_R,\delta \mathcal{Q}_\mathcal{E},\delta \mathcal{Q}_\mathcal{B},\delta \varPi_\parallel,\delta \varPi_{R\mathcal{E}},\delta \varPi_{R\mathcal{B}},\delta \varPi_\mathcal{E},\delta \varPi_\mathcal{B} \}$ contribute to the entropy increase. 

The GS formalism is gauge-invariant if we input the source of the EFE in the form of Eq. \eqref{eq:stress_tensor_decomposition}. However, we need to know the gauge-invariant form of each of the scalar quantities that characterizes the matter perturbations. {The gauge-invariant perturbation $\tilde{\delta X}$ for the quantities $\delta X = \{\delta \varepsilon,\delta s, \delta p, \delta \Pi, \delta \varPi_\parallel\}$ have the general form
\begin{align}
    \delta \tilde X = \delta X - \epsilon^a \bar X_{,a}, \label{eq:scalar_gi}
\end{align}
with $\epsilon^a$ as defined in section \eqref{eq:epsilon_pert}. }

The quantities $\{\mathcal{V}_R, \mathcal{V}_\mathcal{E}, \mathcal{V}_\mathcal{B}, \delta \mathcal{Q}_R, \delta \mathcal{Q}_\mathcal{E}, \delta \mathcal{Q}_\mathcal{B} \}$ define the vector perturbations in Eqs.~\eqref{eq:vel_pert} and \eqref{eq:heat_flux_pert}. Their gauge-invariant equivalent are
\begin{subequations}
    \begin{align}
        \tilde{\mathcal{V}}_R &= \mathcal{V}_R - u_{a;b}n^a \epsilon^b + \frac{1}{2} \left( u^a n^b - n^a u^b \right) \epsilon_{a;b}, \nonumber \\
        \tilde{\mathcal{V}}_\mathcal{E} &= \mathcal{V}_\mathcal{E} - u_a \epsilon^a, \quad \tilde{\mathcal{V}}_\mathcal{B} = \mathcal{V}_\mathcal{B} , \\
        \delta \tilde{\mathcal{Q}}_R &= \delta \mathcal{Q}_R - \epsilon^a \mathcal{Q}_{,a} - \mathcal{Q} n^a n^b \epsilon_{a;b}, \nonumber \\
        \delta \tilde{\mathcal{Q}}_\mathcal{E} &= \delta \mathcal{Q}_\mathcal{E} -  n_a \epsilon^a\mathcal{Q}, \quad 
        \delta \tilde{\mathcal{Q}}_\mathcal{B} = \delta \mathcal{Q}_\mathcal{B} . 
    \end{align} \label{eq:vector_gi}
\end{subequations}

The gauge-invariant equivalent to the quantities $\{\delta \varPi_{R\mathcal{E}}, \delta\varPi_\mathcal{E}, \delta \varPi_{R\mathcal{B}},\delta \varPi_\mathcal{B}\}$ from the tensor perturbations, are
\begin{align}
    \delta \tilde{\varPi}_{R\mathcal{E}} &= \delta\varPi_{R\mathcal{E}} + \frac{1}{2}r^2 \varPi_\perp n^a G_{,a} , \nonumber \\
    \delta\tilde{\varPi}_{R\mathcal{B}} &= \delta\varPi_{R\mathcal{B}} + \frac{1}{2}r^2 \varPi_\perp n^a K_{,a}, \nonumber \\
    \delta\tilde{\varPi}_\mathcal{E} &= \delta\varPi_\mathcal{E} - \varPi_\perp G,  \nonumber \\
    \delta\tilde{\varPi}_\mathcal{B} &= \delta\varPi_\mathcal{B} - \varPi_\perp K. \label{eq:tensor_gi}
\end{align}
The gauge-invariant perturbations of Eqs.~\eqref{eq:scalar_gi}, \eqref{eq:vector_gi}, and \eqref{eq:tensor_gi} are equal to the perturbations in the RW gauge ($\epsilon_a = G = K=0$). In such gauge, the scalar quantities that define the fluid perturbations are equivalent to their gauge-invariant counterpart.

Therefore, the perturbation of the stress tensor in the GS form of Eq. \eqref{eq:stress_tensor_decomposition} from the non-perfect fluid in the RW gauge is 
\begin{subequations}
    \label{eq:stress_tensor_pert_GS_polar}
 \begin{align}
    Q_{ab} =& \left(\delta \varepsilon + \varepsilon \left( \eta - \phi \right) + \mathcal{Q}\left( 2 \mathcal{V}_R + \psi\right)  \right)u_a u_b \nonumber\\
    &+ \left( \delta \mathcal{P}_\parallel + \mathcal{P}_\parallel \left( \eta + \phi \right) + \mathcal{Q} \left( 2 \mathcal{V}_R - \psi \right) \right) n_a n_b \nonumber \\
    &+ \bigg( \delta\mathcal{Q}_R + \frac{1}{2}\mathcal{Q} \left( \eta - \phi \right)  +  \varepsilon \left( \mathcal{V}_R - \frac{\psi}{2}\right) \nonumber \\ & \qquad + \mathcal{P}_\parallel \left( \mathcal{V}_R + \frac{\psi}{2}\right)  \bigg) q_{ab},\\
    Q_a =& \left( \delta \mathcal{Q}_\mathcal{E} + \mathcal{V}_\mathcal{E} \left( \varepsilon + \mathcal{P}_\perp \right)   \right) u_a +  \left( \delta \varPi_{R\mathcal{E}}+\mathcal{V}_\mathcal{E} \mathcal{Q} \right) n_a, \\
    Q =& \delta \mathcal{P}_\perp  + H \mathcal{P}_\perp  , \\
    P =& \delta \varPi_{\mathcal{E}},
\end{align}   
\end{subequations}
for the polar sector.  While, for the axial sector is
\begin{subequations}
\label{eq:stress_tensor_pert_GS_axial}
 \begin{align}
    L_a =& \left( \delta \mathcal{Q}_\mathcal{B} + \mathcal{V}_\mathcal{B} \left( \varepsilon + \mathcal{P}_\perp \right)   \right) u_a \nonumber \\
    &+ \left( \delta \varPi_{R\mathcal{B}} + \mathcal{V}_\mathcal{B} \mathcal{Q}  - \varPi_\perp n_b k^b \right)n_a, \\
    L =& \delta \varPi_{\mathcal{B}},
\end{align}
\end{subequations}
where we have defined $\delta \mathcal{P}_\parallel \equiv \delta \mathcal{P} + \delta \varPi_\parallel$, and $\delta \mathcal{P}_\perp\equiv\delta \mathcal{P} + \delta \varPi_\perp$, as the perturbations of the longitudinal and transverse pressures, respectively. These equations are the source for the perturbed Einstein equations referred in Eqs. \eqref{eq:einstein_pert_polar} and \eqref{eq:einstein_pert_axial}.

\section{Field equations of matter and metric perturbations.} \label{sec:evolution}

We describe the field equations for the metric and matter perturbations of a non-perfect fluid.
The dynamics of spacetime are determined by the perturbed Einstein equations \eqref{eq:einstein_pert_polar} and \eqref{eq:einstein_pert_axial}, using the stress tensor in the form of Eqs. \eqref{eq:stress_tensor_pert_GS_polar} and \eqref{eq:stress_tensor_pert_GS_axial}. These equations yield the evolution and constrains of the metric perturbations. The perturbation to linear order of the conservation equations \eqref{eq:current_cons} and \eqref{eq:stress_cons}, as well as the MIS equations \eqref{eq:Mis_equations}, define the dynamics of the perturbed non-perfect fluid. 
The polar and axial perturbations evolve independently to linear order, and consequently, we study them separately. 

\subsection{Polar perturbations} 

We follow the approach from Gundlach and Martín-García \cite{Gundlach:1999bt}. First, we study the fluid dynamics through the conservation and MIS equations perturbed to linear order. Then, we express the polar Einstein equations \eqref{eq:einstein_pert_polar} as a linear combination of the projection of the tensor basis $\{\Sigma_{ab}, p_{ab}, q_{ab}\}$, and the vector basis $\{u_a, n_a\}$, leading to seven independent equations. Instead of using $\phi$ as a metric perturbation, we introduce $\chi \equiv \phi + \eta - H$. We consider the same linear combination as in \cite{Gundlach:1999bt}, resulting in the perfect fluid equations plus the non-perfect fluid contributions.

The local conservation equations \eqref{eq:current_cons} and \eqref{eq:stress_cons} define the fluid dynamics. We perturb the quantities to linear order, to obtain
\begin{subequations}
    \label{eq:cons_pert}
 \begin{eqnarray}
    (l \geq 0): & \quad 
    \dot{\delta n} + n \left( \mathcal{V}_R + \frac{\psi}{2}\right)' &= \dots, \label{eq:cons_pert_a}\\
    &\dot{\delta \varepsilon}   + \left( \varepsilon + \mathcal{P}_\parallel \right) \left( \mathcal{V}_R + \frac{\psi}{2}\right)' & \nonumber \\
    & + \delta\mathcal{Q}_R' + 2 \mathcal{Q}\left( \dot{\mathcal{V}_R} - \frac{\dot\psi}{2}\right) &= \dots,  \label{eq:cons_pert_b}\\
    &\delta \mathcal{P}_\parallel'  + \left( \varepsilon + \mathcal{P}_\parallel\right) \left( \dot{ \mathcal{V}_R} - \frac{\dot \psi}{2}\right) & \nonumber \\
    &+ \dot{\delta \mathcal{Q}_R} + 2 \mathcal{Q}\left( \mathcal{V}_R + \frac{\psi}{2}\right)'  &= \dots, \label{eq:cons_pert_c}\\
    (l\geq 1): &\quad \left( \varepsilon + \mathcal{P}_\perp \right) \dot{\mathcal{V}_\mathcal{E}} + \dot{\delta \mathcal{Q}_\mathcal{E}} + \delta \varPi_{R\mathcal{E}}' &= \dots, \label{eq:cons_pert_d}
\end{eqnarray}
\end{subequations}
where we have only written the derivatives of the fluid perturbations. The full equations are Eqs. (\ref{eq:matter_polar_complete_a}-{\ref{eq:matter_polar_complete_d}) in the Appendix \ref{sec:complete_eqs}. 
We have calculated the conservation of the fluid current perturbation $\mathbf{\delta J}$, yielding an evolution equation \eqref{eq:cons_pert_a} for $\delta n$. As we work with the energy perturbation $\delta \varepsilon$, and the entropy perturbation $\delta s$, we apply the fundamental law of thermodynamics to the current conservation \eqref{eq:cons_pert_a} to obtain an evolution equation for $\delta s$. 
We also apply the current conservation equation \eqref{eq:currentcons}, the energy equation \eqref{eq:conservation} and the Euler equation \eqref{eq:eulercons} in the background.
This yields the conservation equation for the perturbed entropy $\delta s$, \begin{equation}
    (l\geq 0): \quad \dot{\delta s} + \left( \mathcal{V}_R +\frac{\psi}{2}\right) s' = \dots \label{eq:entropy_cons}
\end{equation}
The full expression can be found in Eq. \eqref{eq:matter_polar_complete_e} in the Appendix \ref{sec:complete_eqs}. Notice that the r.h.s. terms vanish when the background is in equilibrium.

The MIS equations \eqref{eq:Mis_equations} determine the evolution of the dissipative terms. In the polar sector, the dissipative perturbations correspond to $\{\delta \Pi,  \delta \mathcal{Q}_R, \delta\mathcal{Q}_\mathcal{E}, \delta \varPi_\parallel , \delta \varPi_{R\mathcal{E}}, \delta \varPi_\mathcal{E}\}$. The evolution of the perturbation of the viscous pressure $\delta \Pi$ ($l\geq 0$) from Eq. \eqref{eq:MIS_bulk} is 
\begin{equation}
    \tau_\zeta \dot{\delta \Pi}+ \delta \Pi = - \frac{1}{2} \zeta \left( \dot \chi + 3 \dot H\right) + \dots \label{eq:bulk_viscosity_pert},
\end{equation}
{where we have only written the comoving time derivative terms on the metric perturbations}. 
The complete {expression, including the terms omitted}, is in Eq. \eqref{eq:bulk_complete} in the Appendix \ref{sec:complete_eqs}. 

The evolution of the heat flux perturbations $\delta \mathcal{Q}_R$ ($l\geq~0$) and $\delta \mathcal{Q}_\mathcal{E}$ ($l\geq~1$) are obtained by projecting $n^a$ to the time-radial component and the polar part of the angular component of Eq. \eqref{eq:MIS_heat}, given by Eqs. \eqref{eq:radial_heat_complete} and \eqref{eq:polar_heat_complete} respectively, 
\begin{subequations}
\begin{align}
    \tau_\kappa \dot{\delta \mathcal{Q}_R} + \delta \mathcal{Q}_R = & - \kappa T \left( \frac{\delta T}{T} \nu + \dot{\mathcal{V}_R}  \right) + \dots, \label{eq:heat_radial_pert}\\ 
    \tau_\kappa \dot{\delta \mathcal{Q}_\mathcal{E}} + \left( 1 - U \tau_\kappa \right)\delta \mathcal{Q}_\mathcal{E} = & - \kappa T \left( \frac{\delta T}{T} +  \dot{\mathcal{V}_\mathcal{E}} \right) + \dots \label{eq:heat_polar_pert}
\end{align}
\end{subequations}
The evolution of the heat flux perturbations is determined by the acceleration from the velocity perturbations, $\dot{\mathcal{V}_R}$ and $\dot{\mathcal{V}_\mathcal{E}}$, and the temperature perturbations.
The temperature perturbation $\delta T$ is expressed as a function of the independent equilibrium variables using the EoS, similarly to the pressure perturbation in Eq. \eqref{eq:pert_pressure}.

Finally, the polar perturbation of the anisotropic stress tensor has three degrees of freedom: $\delta \varPi_\parallel$ ($l\geq 0$), $\delta \varPi_{R\mathcal{E}}$ ($l\geq 1$), and $\delta\varPi_\mathcal{E}$ ($l \geq 2$). Their evolution equations are obtained by projecting $n^a n^b$ into the time-radial components, the polar mixed components and traceless polar part of the angular components of the MIS equation, respectively. The full expressions for the perturbed anisotropic stress can be found in Eqs.(\ref{eq:anisotropic_complete}-~\ref{eq:anisotropic_E}),
\begin{subequations}
    \begin{align}
        \tau_\lambda \dot{\delta \varPi_\parallel} + \delta \varPi_\parallel &= -\frac{2}{3}\lambda \dot \chi + \dots, \label{eq:anis_viscosity_pert} \\ 
        \tau_\lambda \dot{\delta \varPi_{R\mathcal{E}}} + \left( 1 - U \tau_\lambda \right) \delta \varPi_{R\mathcal{E}} &= - \lambda \mathcal{V}_\mathcal{E}'  + \dots, \label{eq:anis_viscosity_RE}\\ 
        \tau_\lambda \dot{\delta \varPi_\mathcal{E}} + \left( 1 - 2U \tau_\lambda\right) \delta \varPi_\mathcal{E} &= - 2\lambda \mathcal{V}_\mathcal{E} . \label{eq:angular_stress_polar}
    \end{align}        
\end{subequations}
We find that the evolution of the anisotropic stress $\delta \varPi_\parallel$ is related to the evolution of the metric perturbation $\chi$.
However, the evolution of the radial-angular $\delta \varPi_{R\mathcal{E}}$ and pure angular $\delta \varPi_\mathcal{E}$ anisotropic stress is constrained by a spatial derivative and the value of the angular velocity $\mathcal{V}_\mathcal{E}$, respectively. Therefore, their evolutions are dependent on the value of velocity and metric perturbation on each time slice.

The equations for energy conservation \eqref{eq:cons_pert_b} and entropy conservation \eqref{eq:entropy_cons} determine the evolution of the thermodynamic perturbations. The {Euler} equations for radial \eqref{eq:cons_pert_c} and angular {displacement}  \eqref{eq:cons_pert_d} evolve the velocity perturbations. Finally, the MIS equations (\ref{eq:bulk_viscosity_pert}, \ref{eq:heat_radial_pert}, \ref{eq:anis_viscosity_pert}, \ref{eq:anis_viscosity_RE}, \ref{eq:angular_stress_polar}) describe the dynamics of the dissipative terms. Therefore, these equations evolve the matter perturbations for a non-perfect fluid.

{The evolution of the metric perturbations is determined by the EFE \eqref{eq:einstein_pert_polar}. We distinguish two cases: the radiative multipoles, $l\geq 2$, which have gauge-invariant perturbations, and the lower-order multipoles, $l=0,1$, that are not gauge-invariant.}

\subsubsection{Radiative multipoles $l\geq 2$}
We start with the dynamics of the perturbed spacetime for the radiative multipoles, $l\geq 2$. 
The EFE for the polar sector in Eq. \eqref{eq:einstein_pert_polar} are seven independent equations that describes the metric perturbations $\chi$, $\eta$, $\psi$, and $H$, sourced by the stress-tensor perturbations in the form of Eq. \eqref{eq:stress_tensor_pert_GS_polar}.
We use the same set of linear combinations of the EFE as in \cite{Gundlach:1999bt}, in order to recover the perfect fluid results.

The linear combinations of r.h.s. of the EFE \eqref{eq:einstein_pert_polar},
 \begin{align*}
    n^a n^b Q_{ab} - Q - 2\left( (n^aQ_a)' + (2\nu - W) n^a Q_a\right) , \\
    n^an^b Q_{ab} - c_s^2u^a u^b Q_{ab} + 4 W n^a Q_a, \\
    n^a Q_a,
\end{align*}
where $c_s^2$ is the sound speed as used in Eq. \eqref{eq:pert_pressure}, lead to the Einstein equations for the evolution of the metric perturbations,
\begin{align}
    - \ddot \chi + \chi''  =& -16 \pi \Big[ \frac{3}{2} \delta \varPi_\parallel +  \frac{3}{2}\varPi_\parallel \left( 2 \eta - H - \chi\right)  \nonumber \\ 
    & \phantom{-16 \pi \Big[ }+ 2 \mathcal{Q} \mathcal{V}_R - 2 (\delta \varPi_{R\mathcal{E}} + \mathcal{V}_\mathcal{E} \mathcal{Q})' . \nonumber \\ 
    & \phantom{-16 \pi \Big[} +  2(W - 2\nu ) \left(\delta \varPi_{R\mathcal{E}} + \mathcal{V}_\mathcal{E} \mathcal{Q}\right) \Big] \nonumber \\ 
    &+ 2 \left( \mu -U \right)\psi'+ \dots, \label{eq:wave_chi} \\
    - \ddot H + c_s^2 H''  =& 8\pi \Big[ \delta \Pi + \delta \varPi_\parallel + \left( \Pi + \varPi_\parallel\right)\left( 2 \eta -\chi - H \right) \nonumber \\ 
    & \phantom{8\pi [} +2 \mathcal{Q} \mathcal{V}_R (1 - c_s^2)+ \mathcal{Q}\psi\left( 1 + c_s^2 \right) \nonumber \\ 
    &\phantom{8\pi [} +4 W \left( \delta \varPi_{R\mathcal{E}} + \mathcal{V}_\mathcal{E} \mathcal{Q}\right)     \Big] \nonumber \\ 
    &+ 2c_s^2 U \psi' + \dots, \\
    \dot \psi =& 16 \pi \left( \delta \varPi_{R\mathcal{E}} + \mathcal{V}_\mathcal{E} \mathcal{Q}\right) + \dots \label{eq:dot_psi}
\end{align}

The next three combinations of r.h.s. of the EFE,
\begin{align*}
    q_{ab}Q^{ab}, \\
    u_a u_b Q^{ab}, \\
    u_a Q^a,
\end{align*}
yield the constraint equations for the radial derivatives of the metric perturbations, 
\begin{align}
    \left( \dot H \right)' &= 8 \pi \Big[ \left(\delta \Pi +\delta \varPi_\parallel\right) \left( \mathcal{V}_R - \frac{\psi}{2}\right)  \nonumber \\ 
    & \phantom{8 \pi [} + \delta \mathcal{Q}_R + \frac{1}{2}\mathcal{Q}\left( 2\eta - \chi - H \right) \Big] + \dots, \\ 
    H'' &= - 16 \pi \mathcal{Q} \mathcal{V}_R + \dots, \\ 
    \psi' &= 16 \pi \left( \delta \mathcal{Q}_\mathcal{E} + \left( \Pi - \frac{1}{2}\varPi_\parallel \right) \mathcal{V}_\mathcal{E} \right) + \dots \label{eq:prime_psi}
\end{align}

We have only written explicitly the terms from the dissipative contributions of the fluid. Due to the linearity of the EFE, the dissipative {contribution to the source terms are added to the perfect fluid terms}. %
However, the evolution of the dissipative terms is coupled to the perfect fluid terms, as they are coupled in the fluid dynamics equations. The full expressions can be found in Eqs. (\ref{eq:einstein_polar_complete_a}-~\ref{eq:einstein_polar_complete_f}) in the Appendix \ref{sec:complete_eqs}. The equations \eqref{eq:dot_psi} and \eqref{eq:prime_psi} with the derivatives of the metric perturbation $\psi$, depend only on the matter perturbation and the rest of metric perturbations and their derivatives. Therefore, they may be used to simplify the perturbed matter conservation equations \eqref{eq:cons_pert}. 

Finally, the metric perturbation $\eta$ is set by the last Einstein equation \eqref{eq:einstein_pert_polar_d},
\begin{equation}
    \eta = - 8 \pi \delta \varPi_\mathcal{E}, \label{eq:einstein_eta_pi}
\end{equation}
Then, the perturbation on the angular anisotropic stress fixes the trace of the metric perturbation $h_{ab}$, for $l \geq 2$.

We have a set of second order partial differential equations, such that the evolution of the matter and metric perturbations are strongly coupled.  In order to simplify the system {of equations}, we reduce the second order to first order differential equation by introducing $\dot \chi$, $\chi'$, $\dot H$ and $H'$ as variables. Using the evolution and constrain equations for the metric perturbation $\psi$ (\ref{eq:dot_psi},\ref{eq:prime_psi}), we eliminate its derivatives. Thus, seven metric perturbations $\{\psi, \chi, \dot \chi, \chi',H, \dot H, H'\}$ and ten matter perturbation $\{\delta \varepsilon, \delta s, \delta \Pi, \delta \varPi_\parallel, \delta \varPi_{R\mathcal{E}}, \delta \varPi_{\mathcal{E}}, \mathcal{V}_R, \mathcal{V}_\mathcal{E}, \delta \mathcal{Q}_R, \delta \mathcal{Q}_\mathcal{E}\}$ describe the system. Then, we have 17 evolution equations for all the perturbations, and 5 constrain equations, therefore 12 functions are chosen freely on a Cauchy surface to evolve the system. 

Even with the reduction to first order differential equations, this system is still difficult to solve due to the coupling between equation and the setting of initial data.
In a practical scenario, one needs to make certain assumptions on the heat or anisotropic effects. However, we comment two cases: the perfect fluid and the background in equilibrium.

The perfect fluid case on a time dependent background was investigated by Seidel in \cite{Seidel:1990xb}. They found that the choice of $\{\delta s,\chi, H, \psi,  \dot \chi, \dot H\}$, as the free function on the Cauchy surface, evolves such perturbations independently of the rest of matter perturbations. 
In such case, the perturbation $\chi$ {is subjected to a wave equation}, that evolves independently of matter perturbations. The $H$ metric perturbation behaves as a sound wave with velocity $c_s$ or a longitudinal gravitational wave. It is just sourced by the matter perturbations in the form of entropy perturbations $\delta s$ \cite{Gundlach:1999bt}.  

When the background is in equilibrium, the background quantities are set by Eq. \eqref{eq:equilibrium_bg} in the evolution and constrain equations. At first sight, the evolution equations for $\chi$, $H$ and $\psi$ are only sourced by the {viscous} stress quantities $\{\delta \Pi, \delta \varPi_\parallel, \delta \varPi_{R\mathcal{E}},\delta \varPi_\mathcal{E}\}$, and the entropy perturbation $\delta s$. A choice would be to select them and the metric perturbation as the free quantities to evolve independently of the rest of perturbations. However, the evolution of the {viscous} stress is determined by the MIS equations, which depends on the evolution of the rest of the matter perturbation such as $\delta \varepsilon$, the velocities $\mathcal{V}_R$, $\mathcal{V}_\mathcal{E}$ and the heat fluxes $\delta \mathcal{Q}_R$ and $\delta \mathcal{Q}_\mathcal{E}$. The evolution of the system gets simplified when the background is in equilibrium, but the matter and metric perturbations are still coupled.

\subsubsection{Lower-order multipoles l=0,1.} \label{sec:polar_low}
We have focused on the radiative {multipoles} $l \geq 2$, but for the sake of completeness we also mention the non-radiative or lower-order multipoles $l=0,1$. These multipoles need to be studied separately, as the metric perturbations studied from last sections are not gauge-invariant in these cases. 
These cases have two and one gauge degrees of freedom, respectively. Therefore, some gauge choices can be done to simplify the evolution and constrain equations presented in this section. For a discussion on choosing a proper gauge, see \cite{Gundlach:1999bt}.

In the $l=0$ case, scalar quantities resulting from vector and tensor perturbations vanish. Therefore, the only defined metric and matter perturbations are,
\begin{subequations}
    \begin{align}
    \delta \Sigma_{ab} &= h_{ab}^{00} Y_{00}, \\
    \delta \gamma_{AB} &= r^2 \Omega_{AB} H^{00}Y_{00},\\
    \delta T_{ab} &= Q_{ab}^{00} Y_{00}, \\ 
    \delta T_{AB} &= r^2 \Omega_{AB} Q^{00}Y_{00}.
    \end{align}
\end{subequations}
The Einstein equations (\ref{eq:einstein_pert_polar_a},~\ref{eq:einstein_pert_polar_b}) determine the evolution of the metric perturbations. However, the equations (\ref{eq:einstein_pert_polar_c},~\ref{eq:einstein_pert_polar_d}) are not defined for $l= 0$.  The scalar metric decomposition into $\eta$, $\chi$, $\psi$ and $H$ is still valid, then four independent equations are obtained using the tensor basis \eqref{eq:tensorbasis} into \eqref{eq:einstein_pert_polar_a} and \eqref{eq:einstein_pert_polar_b}.

The only scalar fields defined $\{\delta \varepsilon,\delta s,\delta \Pi,\delta \mathcal{Q}_R,\delta \varPi_\parallel \}$ characterize the matter perturbations. The evolution of matter follows the conservation equations (\ref{eq:cons_pert_a}-~\ref{eq:cons_pert_c}). The evolution of the dissipative terms are defined for the  viscous pressure $\delta \Pi$ in Eq. \eqref{eq:bulk_viscosity_pert}, heat flux in the radial direction $\delta \mathcal{Q}_R$ in Eq. \eqref{eq:heat_radial_pert}, and the anisotropic stress $\delta \varPi_\parallel$ in Eq. \eqref{eq:anis_viscosity_pert}, where the non defined perturbations are to be set to zero. The case $l=0$, corresponds to a perturbation on the mass of the system in the vacuum case \cite{Zerilli:1970se}. Then, this perturbation can be removed with a redefinition of the background metric.   

In the $l=1$ case, the metric and matter perturbations are
\begin{subequations}
    \begin{align}
        \delta \Sigma_{ab} &= h_{ab}^{1m} Y_{1m}, \\
        \eta_{aA} &= h_a^{1m} \big[ V^{\mathcal{E}}_{1m}\big]_A, \\
        \delta \gamma_{AB} &= r^2 \Omega_{AB} H^{1m}Y_{1m}, \\
        \delta T_{ab} &= Q_{ab}^{1m} Y_{1m}, \\ 
        \delta T_{aA} &= Q_a^{1m} \big[ V^{\mathcal{E}}_{1m}\big]_A, \\
        \delta T_{AB} &= r^2 \Omega_{AB} Q^{1m}Y_{1m},
    \end{align}   
\end{subequations}
and the rest of the fields are not defined. For such case, the only Einstein equation that is not defined is \eqref{eq:einstein_eta_pi}. 
The rest of Einstein equations from the radiative case are well-defined by setting $l=1$ in Eqs. (\ref{eq:wave_chi}-\ref{eq:prime_psi}).

The matter evolution equations \eqref{eq:cons_pert} are valid for any $l\geq1$. The evolution of the dissipative term are the Eqs. (\ref{eq:bulk_viscosity_pert},~\ref{eq:heat_radial_pert},\ref{eq:heat_polar_pert},~\ref{eq:anis_viscosity_pert},~\ref{eq:anis_viscosity_RE}) by setting $\delta \varPi_\mathcal{E} =0$. In vacuum, these modes are related to the center of mass of the object, and are removed by a proper coordinate change \cite{Zerilli:1970se}.

\subsection{Axial perturbations} 
The axial sector is simpler than the polar one due to the lesser number of variables. The axial sector has one conservation equation for the fluid \cite{Gerlach:1979rw}, 
\begin{equation}
    \nabla^a \left( r^2 L_a \right) = (l-1)(l+2)L,
\end{equation}
which in our fluid description resolves to,
\begin{equation}
    \left( \varepsilon + \mathcal{P}_\perp \right) \dot{\mathcal{V}_\mathcal{B}} + \dot{ \delta \mathcal{Q}_\mathcal{B}} + \mathcal{Q} \mathcal{V}_\mathcal{B}' = \dots \label{eq:axial_conservation}
\end{equation}
The full equation \eqref{eq:axial_evolution} is in the Appendix \ref{sec:complete_eqs}. This expression describes the evolution of the velocity and heat perturbations in the axial sector. 

There are only three dissipative {matter} perturbations, then, only three evolution equation emerge from the perturbed MIS equations \eqref{eq:Mis_equations}. 
The viscous pressure perturbation is purely polar, so there is no axial equivalent. The heat flux perturbation $\delta \mathcal{Q}_\mathcal{B}$ ($l\geq 1$) only correspond to the axial angular component of the MIS heat equation \eqref{eq:MIS_heat}, yielding,
\begin{equation}
    \tau_\kappa \dot{\delta \mathcal{Q}_\mathcal{B}} + \left(1 - U \tau_\kappa\right) \delta \mathcal{Q}_\mathcal{B} = - \lambda T \dot{\mathcal{V}_\mathcal{B}} + \dots \label{eq:axial_heat}
\end{equation}
This evolution equation relates both velocity $\mathcal{V}_\mathcal{E}$ and heat perturbations $\delta \mathcal{Q}_\mathcal{E}$. The equilibrium case implies that $\dot{\mathcal{V}_\mathcal{B}} =0$, which is the same as obtained setting the dissipative quantities to zero in Eq. \eqref{eq:axial_conservation}.

The evolution of the anisotropic stress on the axial sector is determined by the radial-angular and angular components of the MIS equation. The anisotropic stress MIS equation for $\delta \varPi_{R\mathcal{B}}$ ($l \geq 1$) and $\delta \varPi_\mathcal{B}$ ($l\geq 2$) are
\begin{subequations}
 \begin{align}
    \tau_\zeta \dot{\delta \varPi_{R\mathcal{B}}} + \left( 1 - U \tau_\zeta \right) \delta \varPi_{R\mathcal{B}} &= - \lambda \mathcal{V}_\mathcal{B}' + \dots, \label{eq:axial_anis_RB}\\ 
    \tau_\zeta \dot{\delta \varPi_\mathcal{B}} + \left( 1 - 2U \tau_\zeta \right) \delta \varPi_\mathcal{E} &= - 2 \lambda \left( \mathcal{V}_\mathcal{B}  - k_a u^a \right) \label{eq:axial_anis_B}
\end{align}
\end{subequations}
In the axial sector, we find that the evolution of the anisotropic stresses is constrained by the axial velocity perturbation $\mathcal{V}_\mathcal{B}$, and its spatial derivative, $\dot{\mathcal{V}_\mathcal{B}}$, as well as the projection of the metric perturbation $k_a u^a$. As a consequence, the evolution is dependent on the value of velocity and metric perturbation on each time slice.

The conservation equation \eqref{eq:axial_conservation} describes the evolution of the axial velocity perturbation $\mathcal{V}_\mathcal{B}$. 
The MIS equations (\ref{eq:axial_heat}, \ref{eq:axial_anis_RB}, \ref{eq:axial_anis_B}) define the evolution of the dissipative terms {$\delta \mathcal{Q}_\mathcal{B}$, $\delta \varPi_{R \mathcal{B}}$, and $\delta \varPi_{\mathcal{B}}$}. These coupled equations completely characterize the dynamics of the fluid perturbations.

\subsubsection{Axial perturbations $l\geq 2$.} \label{sec:axial_low}
Gerlach and Sengupta \cite{Gerlach:1979rw} showed that the Einstein equations for the axial sector \eqref{eq:einstein_pert_axial} is expressed into one wave equation in the $l \geq 2$ case,
\begin{equation}
    \nabla^a \left(r^{-2} \nabla_a \left( r^4 \Psi \right) \right) - (l-1)(l+2) \Psi = - 16 \pi \epsilon^{ab} L_{a;b}, \label{eq:axial_wave} 
\end{equation}
where $\Psi \equiv \epsilon^{ab} \nabla_b\left( r^{-2} k_a \right)$ is a scalar quantity that characterizes the evolution of the axial metric perturbation.
In our description, the source term in the wave equation is proportional to
\begin{align}
     \epsilon^{ab} L_{a;b} & = \left( \varepsilon + \mathcal{P}_\perp \right) \mathcal{V}_\mathcal{B}' + \delta \mathcal{Q}_\mathcal{B}' + \mathcal{Q}\dot{\mathcal{V}_\mathcal{B}} \nonumber \\
     &+ \dot{\delta\varPi_{R\mathcal{B}}} + \nu \delta \mathcal{Q}_\mathcal{B} + \mu \delta \varPi_{R\mathcal{B}} + \frac{1}{2} k_a n^a \dot{\varPi_\parallel}\nonumber \\
     & + \mathcal{V}_\mathcal{B}\left( \varepsilon' + \mathcal{P}_\perp' + \nu \left( \varepsilon + \mathcal{P}_\perp \right) + \dot{\mathcal{Q}} + \mu \mathcal{Q}\right) \nonumber \\
     & + \frac{1}{2} \varPi_\parallel \left( n^a n^b k_{a;b} + \mu k_a n^a + \nu k_a u^a \right).  \label{eq:axial_source} 
\end{align}  

In the axial sector, we have a wave equation for a scalar degree of freedom, $\Psi$, sourced by the matter perturbations. As in the polar sector, the evolution of all the perturbation are highly coupled. However, the lesser number of variables causes a simpler system of differential equations. We comment the same two cases as in the polar sector: the perfect fluid and the fluid in equilibrium.

In the case of a perfect fluid, the evolution of the metric perturbation $\Psi$, is independent of the metric perturbations as it is {only} sourced by $\mathcal{V}_\mathcal{B}$. In addition, both perturbations evolve independently of each other \cite{Gundlach:1999bt}, {as the axial velocity is constant along the fluid worldlines, $\dot{\mathcal{V}_\mathcal{B}}=0$}. If the background fluid is in equilibrium, the Eqs. \eqref{eq:equilibrium_bg} set the background quantities, and the source in Eq. \eqref{eq:axial_source} turns into
\begin{equation*}
    (\varepsilon + \mathcal{P})\mathcal{V}_\mathcal{B}' + \mathcal{V}_\mathcal{B}\varepsilon' + \delta \mathcal{Q}_\mathcal{B}' + \dot{\delta \varPi_{R\mathcal{B}}} + \nu \delta \mathcal{Q}_\mathcal{B}.
\end{equation*}
Due to the causal form of the MIS equation, the evolution of the matter perturbations is strongly coupled with the evolution of $\Psi$. {Then, similarly to the polar sector, the system of equations is difficult to solve even when the background is in equilibrium.}

\subsubsection{Low multipole perturbations $l=1$.}

For the dipole case $l=1$, the matter perturbation $L$ is not defined. However, the conservation equation \eqref{eq:axial_conservation} is still well-defined for this case. As showed by Gerlach-Sengupta \cite{Gerlach:1979rw} the scalar function $\Psi$ is a solution of a scalar $Z$ as
\begin{align}
    r^4 \Psi = 16 \pi Z + C, \label{eq:axial_l1}
\end{align}
with $C$ an integration constant that can be regularized through boundary conditions. The scalar $Z$ is obtained through integration from the expression of $L_a$, 
\begin{subequations}
 \begin{align}
    Z' &= -r^2 \left( \mathcal{V}_\mathcal{B} \left( \mathcal{E} + \mathcal{P}_\perp\right) + \delta\mathcal{Q}_\mathcal{B}\right) , \\ 
    \dot Z &= r^2 \left( \mathcal{V}_\mathcal{B} \mathcal{Q} + \delta \varPi_{R\mathcal{B}} +\frac{1}{2} \varPi_\parallel n_b k^b\right).
\end{align}
\end{subequations}
In the case of perfect fluid, we have $\dot Z=0$, therefore $Z$ is constant along the fluid worldines, implying that $\Psi$ is time independent. However, when we add the dissipative terms, the axial metric perturbation is dependent on time. For the background in equilibrium case, the comoving time derivative of $Z$ depends only on the anisotropic stress as, $\dot Z = r^2 \delta \varPi_{R\mathcal{B}}$.

The equation \eqref{eq:axial_l1} implies that there is no degree of freedom for the axial perturbations as they are constrained by the matter perturbations for the dipole.
Even in such case, $\Psi$ is still gauge-invariant, however the reconstruction of $k_a$ is degenerate as it is built from the gradient of a scalar.

\section{Conclusions and future work}
\label{conclude}
In this work, we have calculated the gauge-invariant perturbations for a self-gravitating relativistic non-perfect fluid in a spherical background. 
We have presented a formalism to treat the dynamics of the non-perfect fluid, including the evolution of the dissipative terms, using extended irreversible thermodynamics.
We have included these terms to the evolution and constrain equations of the gauge-invariant non-spherical metric perturbations from a time-dependent spherical system.

We have used the Gerlach-Sengupta formalism \cite{Gerlach:1979rw,Gerlach:1980tx} to study the non-spherical perturbations for a non-perfect fluid in a gauge-invariant framework. 
We have approached the problem as Gundlach and Martín-García \cite{Gundlach:1999bt}, using the natural projection of the quantities into the fluid velocity $u^a$, and its orthogonal projection $n^a$. We have obtained a set of scalar equations with frame derivatives. All of these quantities are covariant and gauge-invariant, while maintaining a time-dependent background. We used this property to a non-perfect fluid that can be out-of-equilibrium and, by definition, is time-dependent.
We described the equilibrium state of the fluid as Israel \cite{Israel:1979wp} to {define} a set of gauge-invariant dissipative quantities.
In addition, we presented the Müller-Israel-Stewart equations \cite{Muller:1967zza,Israel:1979wp} for the evolution of these terms in this formalism for the first time. Finally, we have calculated a set of evolution and constrain equations for the fluid and metric perturbations in the polar and axial sector, for $l\geq 2$. We have commented the lower-order multipole cases {($l=0,1$)}, for both sectors, including these dissipative terms.

In the polar sector, we have found that the evolution equation for the matter and metric perturbations are strongly coupled. 
The anisotropic stress $\varPi_{\mu \nu}$ sources the evolution of the transverse, $\chi$, and longitudinal gravitational waves, $H$, in the interior of the object. 
Due to the coupling, the heat flux $q_{\mu}$ also influences the evolution of such terms. The trace of the metric perturbation in the time-radial part is sourced by the angular anisotropy $\delta \varPi_\mathcal{E}$. 
In the axial sector, there is only one degree of freedom for the gravitational perturbations $\Psi$. 
The existence of dissipative terms couples the evolution of the matter perturbation with the wave equation of $\Psi$, however, the coupling is less complex than in the polar sector.

In this work, we have not made any assumption on the strength of the viscous effects.
However, The MIS equations only work for slightly out-of-equilibrium conditions. In addition, we have used the Maxwell-Cattaneo form, which makes further simplifications. 
The objective has been to present the extended irreversible thermodynamics constitutive equations in the GS formalism, so we have selected the simplest version of such equations. However, other phenomenological equations may be chosen for the evolution of the dissipative term, as well as completely effective out-of-equilibrium fluid treatment.
For a more rigorous approach, a 2-parameter perturbative formalism \cite{Sopuerta:2003rg} would be fitting, keeping track of the viscous effects and the non-spherical perturbations separately. Our framework is able to work as such to linear order. First, we consider that the background is in equilibrium, then the linear order perturbations  include {perfect fluid and dissipative effects as perturbations from equilibrium}. 
With a multiparameter perturbative approach, one can also treat the radial perturbations, and their coupling to higher order \cite{Passamonti:2004je}.

In conclusion, the presented formalism of the dynamics of a perturbed non-perfect fluid coupled to General Relativity, including viscosity and heat flux, provides a tool to study astrophysical phenomena, where these dissipative effects are important. The framework is particulary well-suited to investigate the stability of stars in scenarios where out-of-equilibrium physics can not be neglected, such as stars mergers or newborn (post-collapse) stars. In such cases, the formalism can be applied to a time-dependent background where the dissipation is important, e.g. neutrino cooling in a newborn star, to study the oscillations modes of any matter perturbation. The covariant and gauge-invariant properties favor the use of any coordinate system where the equations are simpler to solve.

To have a complete picture of the dissipative processes of the star, a description of the energy loss to the exterior of the star through graviational radiation is needed. Such a study of the exterior (vacuum) solution and the matching conditions with the surface of the star would be the natural continuation of this work. This allows the study of the effects of matter dissipation in the incoming gravitational waves, as well as, the damping of the outgoing gravitational waves due to dissipation in the interior of the star.

\begin{acknowledgments}
{D. Díaz-Guerra, C. Albertus and M. A. Pérez-García acknowledge support from Spanish Ministry of Science PID2022-137887NB-10 and RED2022-134411-T projects} and Junta Castilla y Le\'on SA101P24 project. This article is based on collaboration within the COST Action COSMIC WISPers CA21106. P. Char is supported by European Union's HORIZON MSCA-2022-PF-01-01 Programme under Grant Agreement No. 101109652, project ProMatEx-NS.
\end{acknowledgments}


\clearpage
\begin{widetext}
\appendix
\section{Field equations for the perturbed non-perfect fluid} \label{sec:complete_eqs}
Here we express the full equation and constrain equations for the perturbed non-perfect fluid. We use $\chi \equiv \phi + \eta - H$.

\subsection*{Polar sector}~
The perturbed conservation equations for the polar are
\begin{align}
    \dot{\delta n} + n\left( \mathcal{V}_R + \frac{\psi}{2}\right)' =& - \left( \mu + 2 U \right) \delta n - \frac{1}{2}\left( H + \chi - 2\eta \right) \dot{n} - \left( \mathcal{V}_R - \frac{\psi}{2}\right) n'  \nonumber \\
    &- n \left( \frac{1}{2} \left( \dot \chi + 3 \dot H \right) + \frac{1}{2} \left( H + \chi -2\eta \right) \left( \mu + 2U \right) + \left( \mathcal{V}_R + \frac{\psi}{2} \right) \left( \nu + 2W \right) - \frac{l(l+1)}{r^2} \mathcal{V}_\mathcal{E} \right), \label{eq:matter_polar_complete_a}\\
    \dot{\delta \varepsilon}   + \left( \varepsilon + \mathcal{P}_\parallel \right) \left( \mathcal{V}_R + \frac{\psi}{2}\right)' =&  - \mu \left( \delta \varepsilon + \delta \mathcal{P}_\parallel \right) - 2U \left( \delta \varepsilon + \delta \mathcal{P}_\perp \right) - \left( \mathcal{V}_R - \frac{\psi}{2}\right) \varepsilon'   \nonumber \\
    & - \left(  \varepsilon + \mathcal{P}_\parallel \right) \left( \left( \mathcal{V}_R + \frac{\psi}{2} \right) \nu + \frac{1}{2} \left( \dot H + \dot \chi\right) \right) \nonumber \\ 
    &- \left( \varepsilon + \mathcal{P}_\perp\right) \left(\dot H  + 2W \left( \mathcal{V}_R + \frac{\psi}{2}\right) -\frac{l(l+1)}{r^2} \mathcal{V}_\mathcal{E} \right)  \nonumber \\
    & - \delta \mathcal{Q}_R'   - 2 \delta\mathcal{Q}_R\left( \nu + W \right) + \frac{l(l+1)}{r^2} \delta \mathcal{Q}_\mathcal{E}  + \frac{3}{2} \left( H + \chi - \frac{2}{3} \eta \right) \mathcal{Q}' - \left( \mathcal{V}_R + \frac{\psi}{2}\right) \dot{\mathcal{Q}} \nonumber\\
    &- \mathcal{Q} \left( 2 \eta' - \frac{1}{2} ( H' + 3\chi') +  (2\eta - 3 H - 3 \chi) (\nu + W ) + 2 \dot{\mathcal{V}_R} - \dot \psi + 2 \left( \mathcal{V}_R - \frac{\psi}{2} \right) (\mu +U )  \right),\label{eq:matter_polar_complete_b} \\
    \delta \mathcal{P}_\parallel'  + \left( \varepsilon + \mathcal{P}_\parallel\right) \left( \dot{ \mathcal{V}_R} - \frac{\dot \psi}{2}\right) =& -\dot{\delta \mathcal{Q}_R} - 2 \mathcal{Q}\left( \mathcal{V}_R + \frac{\psi}{2}\right)' - \left( \delta \varepsilon + \delta \mathcal{P}_\parallel \right) \nu  - 2 W \left( \delta \mathcal{P}_\parallel - \delta \mathcal{P}_\perp \right) - 2\delta \mathcal{Q}_R \left( \mu + U \right)  +\frac{l(l+1)}{r^2} \delta \varPi_{R\mathcal{E}} \nonumber \\
    &- \left( \varepsilon + \mathcal{P} _\parallel \right) \left( \frac{1}{2} \left( 2\eta'- H' - \chi'\right) + \mu \left( \mathcal{V}_R - \frac{\psi}{2} \right) \right) - \left( \mathcal{P}_\parallel - \mathcal{P}_\perp \right) \left( H' + 2U \left( \mathcal{V}_R - \frac{\psi}{2}\right) \right)  \nonumber \\
   &+ \frac{1}{2}\left( 2 \eta - H - \chi \right) \dot{\mathcal{Q}}  - \mathcal{Q}' \left( \mathcal{V}_R + \frac{\psi}{2}\right)- \left( \mathcal{V}_R - \frac{\psi}{2}\right) \dot{\mathcal{P}_\parallel}  \nonumber \\
    &-\mathcal{Q} \left(\frac{1}{2} \left(\dot \chi + 3\dot H  \right) + (H - \chi - 2\eta)(\mu + U) + \left( 2\mathcal{V}_R + \psi \right) ( \nu + W)  -\frac{l(l+1)}{r^2} \mathcal{V}_\mathcal{E} \right)  , \label{eq:matter_polar_complete_c}\\
    \left( \varepsilon + \mathcal{P}_\perp \right) \dot{\mathcal{V}_\mathcal{E}} + \dot{\delta \mathcal{Q}_\mathcal{E}} =& - \delta\varPi_{R\mathcal{E}}' - \left( \nu + 2 W \right) \delta \varPi_{R\mathcal{E}} - \delta \mathcal{P}_\perp - \delta \mathcal{Q}_\mathcal{E} \left( \mu + 2 U\right) + \frac{(l-1)(l+2)}{2r^2} \delta \varPi_\mathcal{E}\nonumber \\
    & + \frac{1}{2}\left( \eta + H + \chi\right) \varPi_\parallel - \frac{1}{2}\left( 2\eta- H - \chi \right)\left( \varepsilon + \mathcal{P}\right)  - \mathcal{Q}\left( \psi + \mathcal{V}_\mathcal{E}' \right) - \mathcal{V}_\mathcal{E}\left( \mu (\mathcal{P}_\perp - \mathcal{P}_\parallel) - \mathcal{Q}\nu + \dot {\mathcal{P}_\perp} \right), \label{eq:matter_polar_complete_d} \\ 
    n T \left( \dot{\delta s} + \left( \mathcal{V}_R + \frac{\psi}{2}\right) s' \right) &= \left( \mathcal{V}_R - \frac{\psi}{2} \right)\left( p' + \left( \varepsilon + p \right)\nu \right) - \delta \Pi \left( \mu + 2U \right) - \delta \varPi_\parallel \left(\mu - U \right)\nonumber \\ 
    &  + \left( \mathcal{V}_R - \frac{\psi}{2}\right) \Pi' -  \left( \frac{1}{2}\left( \dot \chi + 3 \dot H \right) + \mathcal{V}_R' + \frac{1}{2}\psi'+ 2 W \mathcal{V}_R + \psi ( \nu + W)  - \frac{l(l+1)}{r^2}\mathcal{V}_\mathcal{E} \right) \Pi \nonumber \\ 
    & + \left( \mathcal{V}_R - \frac{\psi}{2}\right) \varPi_\parallel'- \left( \frac{1}{2} \dot \chi + \mathcal{V}_R' + \frac{1}{2}\psi' - 4W\mathcal{V}_R + \psi ( \nu + W)- \frac{l(l+1)}{2r^2}\mathcal{V}_\mathcal{E} \right) \varPi_\parallel \nonumber \\ 
    &- \delta \mathcal{Q}_R' - 2 \delta \mathcal{Q}_R \left(\nu + W\right) + \frac{l(l+1)}{r^2} \delta \mathcal{Q}_\mathcal{E} + \frac{3}{2}\left( H + \chi - \frac{2}{3}\eta \right) \mathcal{Q}' \nonumber \\ 
    &+ \left(\frac{1}{2}\left( H' + 3 \chi' - 4 \eta' \right) + 3 \left(H + \chi - \frac{2}{3}\eta \right) (\nu + W) + \dot \psi - 2 \dot{\mathcal{V}_R}  \right)\mathcal{Q} \label{eq:matter_polar_complete_e}
\end{align}
\newpage

We explicitly write the MIS equations perturbed to first order for the polar sector,~
\begin{align}
    \tau_\zeta \dot{\delta \Pi}+ \delta \Pi =&    - \zeta \left( \frac{1}{2} (\dot \chi + 3 \dot H) + \mathcal{V}_R' + \frac{\psi'}{2} +\left( \nu + 2W\right) \left( \mathcal{V}_R + \frac{\psi}{2}\right) - \frac{l(l+1)}{r^2} \mathcal{V}_\mathcal{E} \right) \nonumber \\
    &+\tau_\zeta \left( \left( \eta - \frac{1}{2}( H + \chi)  \right) \Pi' + \left( \mathcal{V}_R + \frac{\psi}{2}\right) \Pi' \right), \label{eq:bulk_complete}\\ 
    \tau_\kappa \dot{\delta \mathcal{Q}_R} + \delta \mathcal{Q}_R = & - \kappa T \bigg[ \frac{\delta T}{T} \nu + \ln \delta T' + \left( \mathcal{V}_R - \frac{\psi}{2}\right) \dot{\ln T} + \left( \eta - \frac{H}{2} - \frac{\chi}{2}\right) \ln T' \nonumber \\ 
    & +  \left( \dot{\mathcal{V}_R} - \frac{\dot\psi}{2}\right) + \left( \mathcal{V}_R - \frac{\psi}{2}\right) \mu  + \left( \eta' - \frac{H'}{2} - \frac{\chi'}{2}\right) + \left( \eta - \frac{H}{2} - \frac{\chi}{2}\right) \nu \bigg] \nonumber \\ 
    &- \tau_\kappa \left( \left( \mathcal{V}_R + \frac{\psi}{2}\right) \mathcal{Q}'- \frac{1}{2} \left( \dot H + \dot \chi \right) \mathcal{Q}\right) - \left( \eta - \frac{H}{2} - \frac{\chi}{2}\right) \mathcal{Q}, \label{eq:radial_heat_complete} \\ 
    \tau_\kappa \dot{\delta \mathcal{Q}_\mathcal{E}} + \left( 1 - U \tau_\kappa \right) \delta \mathcal{Q}_\mathcal{E} = & - \kappa T \left( \frac{\delta T}{T} +   \dot{\mathcal{V}_E} + \mathcal{V}_\mathcal{E}\dot{\ln T} +\eta - \frac{H}{2} - \frac{\chi}{2} \right) + \tau_\kappa \mathcal{Q}\left( \mathcal{V}_\mathcal{E} \left( \nu - W \right) - \frac{\psi}{2}  \right) \label{eq:polar_heat_complete}\\
    \tau_\lambda \dot{\delta \varPi_\parallel} + \delta \varPi_\parallel =& - \frac{2}{3} \lambda \left( \dot \chi + 2 \mathcal{V}_R ' + \psi' + \left( 2 \mathcal{V}_R + \psi \right)\left( \nu - W \right) + \frac{l(l+1)}{r^2}\mathcal{V}_\mathcal{E}  \right) \nonumber \\  
    & - \tau_\lambda \varPi_\parallel \left( \mathcal{V}_R' + \frac{\psi'}{2}\right) - \varPi_\parallel \left( \eta - \frac{H}{2} - \frac{\chi}{2}\right), \label{eq:anisotropic_complete} \\ 
    \tau_\lambda \dot{\delta \varPi_{R\mathcal{E}}} + \left( 1 - U \tau_\lambda \right) \delta \varPi_{R\mathcal{E}} =& - \lambda\left( \mathcal{V}_\mathcal{E}' + \mathcal{V}_\mathcal{E}\left( \nu - 2 W \right)  + \mathcal{V}_R + \frac{\psi}{2}     \right)  - \frac{3}{2} \tau_\lambda \varPi_\parallel \left( \mathcal{V}_\mathcal{E}( W - \nu) + \frac{\psi}{2} \right), \label{eq:anisotropic_RE}\\ 
    \tau_\lambda \dot{\delta \varPi_\mathcal{E}} + \left( 1 - 2U \tau_\lambda\right) \delta \varPi_\mathcal{E} =& - 2\lambda \mathcal{V}_\mathcal{E}. \label{eq:anisotropic_E}
\end{align}

The Einstein field equations for the polar sector obtained are
\begin{align}
    -\ddot \chi + \chi'' + 2 \left( \mu - U \right) \psi' &=  2 W  \chi'  +3 \mu  \dot \chi + 4 \dot H(\mu - U)  + 2 \dot \eta   (U - \mu) + \eta' (8\nu  -6 W )\nonumber \\
     &+  2 \left( \frac{6m}{r^3} -  8 \pi  \left( \varepsilon +  \mathcal{P}_\perp  - \mathcal{P}_\parallel \right) -2U^2 - 2\nu^2  + 2U \mu  \right)(H + \chi)-  \frac{(l-1)(l+2)}{2r^2}\chi\nonumber\\ 
    &+  \left(-\frac{l (l+1) + 8}{r ^2} +  32 \pi ( \varepsilon + \mathcal{P}_\perp - \mathcal{P}_\parallel ) +4 \left(- U( \mu +2 U) + 2 W( 2W - \nu) +\nu  ^2\right)\right) \eta\nonumber \\
    &+ 2 \psi \left(  2 \mu W  - 2 UW  - \mu' - 2 \mu \nu  + \dot \nu  \right) \nonumber \\
    &- 16 \pi \left( \delta \mathcal{P}_\parallel - \delta \mathcal{P}_\perp  + 2 \mathcal{Q} \mathcal{V}_R - 2\left( (\delta \varPi_{R\mathcal{E}} + \mathcal{V}_\mathcal{E} \mathcal{Q})' + (2\nu - W) \left(\delta \varPi_{R\mathcal{E}} + \mathcal{V}_\mathcal{E} \mathcal{Q}\right)\right) \right), \label{eq:einstein_polar_complete_a} \\
    -\ddot H + c_s^2 H'' - 2c_s^2 U \psi' &=(1 + c_s^2) U \dot \chi - 2 U \dot \eta + \left( 4 U + c_s^2 ( 2U + \mu) \right) \dot H - (1-c_s^2) W \chi' + 2 W \eta' - (\nu + 2c_s^2 W) H' \nonumber \\
    &+ \left(  2 U^2 - \frac{4m}{r^3} + c_s^2\left(\frac{l(l+1)}{r^2} + 2\left( U + 2\mu \right) \right) \right) \left( \chi + H\right) - \frac{(l-1)(l+2)}{2r^2} \chi \nonumber \\ 
    & +\left(  \frac{6m}{r^3} + \frac{l(l+1) - 4}{2r^2} - 3U^2 - c_s^2 U\left(  U + 2 \mu\right) \right) 2\eta +\left((\nu+ W) U(1+c_s^2) - (1-c_s^2)\mu W \right)2\psi \nonumber \\ 
    &+ 8 \pi \left( \delta \mathcal{P}_\parallel-c_s^2 \delta \varepsilon + \mathcal{P}_\parallel\left( 2 \eta -\chi - H \right) - c_s^2 \varepsilon \left( \chi + H \right) +2 \mathcal{Q} \mathcal{V}_R (1 - c_s^2)+ \mathcal{Q}\psi\left( 1 + c_s^2 \right) \right. \nonumber \\
    & \left.+ 4 W \left( \delta \pi_{RE} + \mathcal{V}_\mathcal{E} \mathcal{Q}\right) \right), \label{eq:einstein_polar_complete_b}\\
    \dot \psi &= - \chi' + 2 \eta' - 2 \nu \left( H - \chi \right)  - 2 \mu\psi  - 2\left( \nu - W\right) \eta  + 16 \pi \left( \delta \varPi_{R\mathcal{E}}+\mathcal{V}_\mathcal{E} \mathcal{Q} \right) , \label{eq:einstein_polar_complete_c} \\
    \left(\dot H\right)' &=  \left( \mu - 2U\right) H' + U \left( 2 \eta' - \chi' \right) + W \dot \chi + \left(W (\nu + 2W) - U (\mu + 2U) - \frac{l(l+1) + 2}{2 r^2} \right) \psi   \nonumber \\ 
    &+8 \pi \bigg(  \varepsilon \left( \mathcal{V}_R + \frac{\psi}{2}\right)   + \mathcal{P}_\parallel \left( \mathcal{V}_R - \frac{\psi}{2}\right) + \delta \mathcal{Q}_R + \frac{1}{2}\mathcal{Q} \left(2\eta - \chi - H\right)  \bigg), \label{eq:einstein_polar_complete_d}\\
    H'' &=   W \left( \chi' - 2 H' \right) + U \left( \dot \chi + 2 \psi' \right) +  \left(\mu +2 U \right)\dot H +  \left( \frac{l(l+1)}{r^2} + 4 \mu U + 2 U^2 \right) (H + \chi) \nonumber \\ 
    &- \frac{(l-1)(l+2)}{2r^2} \chi -2U\left( 2 \mu  + U \right) \eta + \left(  2 \mu W + 2U \left( \nu + W \right) \right)\psi \nonumber \\ 
   &- 8 \pi  \left(\delta \varepsilon +\varepsilon\left( H + \chi\right)  + 2 \mathcal{Q} \mathcal{V}_R   \right), \label{eq:einstein_polar_complete_e}\\ 
   - \psi' &= \dot \chi + 2\dot H +2 \nu \psi - 2\left( \mu + U\right) \eta  + 2 H \mu + 2 \mu \chi  -  16 \pi \left( \left( \delta \mathcal{Q}_\mathcal{E} + \mathcal{V}_\mathcal{E} \left( \varepsilon + \mathcal{P}_\perp \right)   \right) \right). \label{eq:einstein_polar_complete_f}
\end{align}

\subsection*{Axial sector}
The perturbed conservation equation for the axial sector is
\begin{align}
    \left( \varepsilon + \mathcal{P}_\perp \right) \dot{\mathcal{V}_\mathcal{B}} + \dot{ \delta \mathcal{Q}_\mathcal{B}} + \mathcal{Q} \mathcal{V}_\mathcal{B}' =& - \delta \varPi_{R \mathcal{B}}' -  \delta \varPi_{R\mathcal{B}} (\nu + 2 W)- \mathcal{V}_\mathcal{B} \left( \dot{\mathcal{P}_\perp} + \mu \left( \mathcal{P}_\perp - \mathcal{P}_\parallel\right) - \nu \mathcal{Q} \right) - \mathcal{Q}_\mathcal{B} (\mu + 2 U)  \nonumber \\
    & + \frac{(l-1)(l+2)}{2r^2} \delta \varPi_\mathcal{B} - \frac{1}{2} \left(k^a n_a  \varPi_\parallel' + \varPi_\parallel \left( n^a n^b k_{a;b} + \nu k^a n_a + 2 W k^a n_a + \mu k^a u_a \right) \right). \label{eq:axial_evolution}
\end{align}

The MIS equations for the axial sector are,
\begin{align}
    \tau_\kappa \dot{\delta \mathcal{Q}_\mathcal{B}} + \left(1 - U \tau_\kappa\right) \delta \mathcal{Q}_\mathcal{B} =& - \kappa T \left( \dot{\mathcal{V}_\mathcal{B}} + \mathcal{V}_\mathcal{B} \dot{\ln T} \right) + \tau_\kappa \mathcal{Q} \left(  \frac{1}{2}r^2 \Psi + \mathcal{V}_\mathcal{B}\left( \nu - W \right)  \right), \\ 
    \tau_\lambda \dot{\delta \varPi_{R\mathcal{B}}} + \left( 1 - U \tau_\lambda \right) \delta \varPi_{R\mathcal{B}} =& - \lambda \left( r^2 \Psi + \mathcal{V}_\mathcal{B}' + \mathcal{V}_\mathcal{B} ( \nu - 2W )+ \frac{2}{3}k_a n^a\left( U - \mu \right) \right) \nonumber \\ 
    & - \frac{3}{2} \tau_\lambda \varPi_\parallel \left(\mathcal{V}_\mathcal{B} ( W - \nu) - \frac{1}{2} r^2 \Psi + 3 \dot{\left( k_a n^a \right)} - 3k_a n^aU  \right), \\ 
    \tau_\lambda \dot{\delta \varPi_\mathcal{B}} + \left( 1 - 2U \tau_\lambda \right) \delta \varPi_\mathcal{B} =& - 2 \lambda \left( \mathcal{V}_\mathcal{B}  - k_a u^a \right).
\end{align}
\end{widetext}

\bibliography{references}

\end{document}